\documentclass[12pt]{article}

\usepackage{amssymb, hyperref, amsmath,amsfonts, bm, eurosym,geometry,ulem,graphicx,caption,mathtools,natbib, xcolor,setspace,comment,footmisc,caption,pdflscape,subfigure,array, parskip, bbm, csquotes, multirow, booktabs,caption}

\hypersetup{hidelinks}
    
\makeatletter
\renewcommand\hyper@natlinkbreak[2]{#1}
\makeatother

\usepackage[flushleft]{threeparttable}

\usepackage[capposition=top]{floatrow}

\usepackage[capposition=top]{floatrow}
\usepackage[toc,page]{appendix}

\graphicspath{ {./images/} }

\newtheorem{theorem}{Theorem}

\newtheorem{proposition}{Proposition}

\newenvironment{proof}[1][Proof]{\noindent\textbf{#1.} }{\hfill$\square$}

\newtheorem{definition}[theorem]{Definition}

\newcolumntype{L}[1]{>{\raggedright\let\newline\\arraybackslash\hspace{0pt}}m{#1}}
\newcolumntype{C}[1]{>{\centering\let\newline\\arraybackslash\hspace{0pt}}m{#1}}
\newcolumntype{R}[1]{>{\raggedleft\let\newline\\arraybackslash\hspace{0pt}}m{#1}}

  {\list{}{\leftmargin=0.3in\rightmargin=0in}\item[]}%
  {\endlist}

\begin{document}

\begin{titlepage}
\title{Competitive equilibrium \\and the double auction\vspace{-0.5em}\thanks{For helpful comments and useful discussions, I would like to thank Jesper Akesson, Miguel Ballester, Adam Brzezinski, David Van Dijcke, Dan Friedman, Steve Gjerstad, Bernhard Kasberger, Erik Kimbrough, John Ledyard, Luke Milsom, Heinrich Nax, Charlie Plott, David Porter and Jasmine Theilgaard. I would also like to thank audiences at Oxford and Michigan State's Quantitative Economics Workshop. Finally, I am grateful to 
Aniket Chakravorty for excellent research assistance and to the George Webb Medley Fund for generous financial support.\vspace{0.2em}}}
\author{Itzhak Rasooly\thanks{Sciences Po, the Paris School of Economics, and the University of Oxford.} }
\date{\today}
\maketitle
\vspace*{-0.4cm}
\begin{center}
\end{center}
\vspace*{-0.8cm}
\begin{abstract}
\noindent In this paper, we revisit the common claim that double auctions necessarily generate competitive equilibria. We begin by observing that competitive equilibrium has some counterintuitive implications: specifically, it predicts that monotone shifts in the value distribution can leave prices unchanged. Using experiments, we then test whether these implications are borne out by the data. We find that in double auctions with stationary value distributions, the resulting prices can be far from competitive equilibria. We also show that the effectiveness of our counterexamples is blunted when traders can leave without replacement as time progresses. Taken together, these findings suggest that the `Marshallian path' is crucial for generating equilibrium prices in double auctions.

\noindent \\
\vspace{-0.75cm}
\vspace{0in}

\bigskip
\end{abstract}
\setcounter{page}{0}
\thispagestyle{empty}
\end{titlepage}
\pagebreak \newpage

\section{Introduction}\label{introduction}

In his `Introduction to Economic Science', \cite{fisher} wrote that `If you want to make a first-class economist, catch a parrot and teach him to say ``supply and demand'' in response to every question you ask him.' Apparently, this joke was considered dated even in 1910 --- it is attributed to a critic of economics writing from a `long time ago' --- and it thus serves to illustrate the dominance of supply and demand in the century before Fisher's publication. However, supply and demand analysis remained popular in Fisher's time; and indeed one of the purposes of Fisher's book was to expound and refine such analysis.

More than a century later, the idea of supply and demand --- or `competitive equilibrium', as we might now call it --- remains pervasive throughout economic theory. Its extension to multiple markets in the form of general equilibrium theory has provided a basis for celebrated welfare theorems \citep{arrow1951extension, debreu1954valuation}, as well as the foundation for much of modern macroeconomics (starting with \cite{lucas1977understanding}, \cite{kydland1982time}, etc.). In addition, partial equilibrium models have been applied to areas as diverse as discrimination \citep{becker1957economics}, marriage \citep{becker1973theory, becker1974theory} and location choice \citep{glaeser2007economics}. Thus, although the notion of competitive equilibrium may not have quite the dominance that it did in Fisher's time --- supplanted in part by the rise of non-cooperative game theory --- it surely remains one of the foundational concepts of economic theorising.

In part, the pervasiveness of competitive equilibrium may be due to the perception that its predictions have been experimentally vindicated by a series of double auction experiments starting with \cite{smith1962experimental}. As \cite{plott1981theories} puts it in an early review of the literature: `The overwhelming result [from these experiments] is that these markets converge to the competitive equilibrium even with very few traders'. In a more recent study of two thousand classroom experiments, \cite{lin2020evidence} reach a similar conclusion, declaring that competitive equilibrium convergence in double auctions appears to be `as close to a culturally universal, highly reproducible outcome as one is likely to get in social science'. They add that such convergence should be considered `as reproducible as the kinds of experiments that are done in a college chemistry laboratory to demonstrate universal chemistry principles'.\footnote{Indeed, existing experimental results appear to be surprisingly robust to changing the number of bidders \citep{smith1965experimental}, changing the cultural context \citep{kachelmeier1992culture}, and even introducing `extreme earnings inequality' at equilibrium \citep{holt1986market, smith2000boundaries, kimbrough2018testing}. Thus, while it may be possible to at least slow convergence to competitive equilibrium through large changes to the underlying environment --- e.g. by allowing for resale as in \cite{dickhaut2012commodity} or market power as in \cite{kimbrough2018testing} --- the existing literature suggests fast convergence to competitive equilibrium in double auction environments similar to those first considered by \cite{smith1962experimental}. See also \cite{ikica2018competitive} for a large-scale replication of competitive equilibrium convergence in standard double auction environments.}

In this paper, we revisit this conclusion. Our starting point is that competitive equilibrium can make highly counterintuitive, but previously unstudied, predictions. To see the basic idea, consider a market with 99 buyers, each with unit demand and with valuations of £1.01, £2.01, £3.01, ..., up to £99.01. Meanwhile, suppose that there are 99 sellers, each possessing just one unit to sell and with valuations (or `costs') of £0.99, £1.99, £2.99, ... up to £98.99. Under such assumptions, one can check that the (essentially unique\footnote{Depending on how one handles ties, prices of £49.99 and £50.01 can also clear the market.}) market-clearing price is £50: at such a price, 50 buyers want to purchase (those with valuations above £50) and 50 sellers want to sell (those with valuations below £50). Imagine now that we decrease the valuations of all sellers whose initial valuation was below £50 by an arbitrary amount, say to £0. Intuitively, one would expect this to drive down the price, both by inducing sellers to accept lower offers and by allowing them to profitably submit lower offers themselves. Despite this, however, the shift does not change the competitive equilibrium price: a price of £50 still generates demand from 50 buyers (since demand has not changed), and still generates supply from the same 50 sellers (who are now even more eager to sell).

In Section \ref{CE preserving shifts}, we begin by generalising the example just given. That is, we identify a broad class of downward shifts to the distributions of buyer and seller valuations which preserve the set of competitive equilibrium prices. As discussed, it is highly counterintuitive that such downward shifts would in fact leave market prices unaltered. As a result, such shifts provide a challenging test for competitive price theory.

We then conduct experimental double auctions to investigate whether such shifts do in fact depress observed prices. One important feature of our initial set of double auction experiments is that they hold the value distributions fixed through use of a queue of buyers and sellers: every time a buyer or seller exits the market, a new buyer or seller takes their place (see \cite{brewer2002behavioral} for a similar methodology). We use this queueing procedure for two reasons. First, it can be justified on grounds of realism: in actual markets, trade does not end after a couple of periods once all willing buyers and sellers have traded. Instead, the market is continually replenished by a new stock of traders. Second, and much more importantly, our design necessarily holds competitive equilibrium fixed, thereby allowing us to rigorously study if double auctions outcomes converge to \textit{the} set of competitive equilibrium prices. In contrast, standard designs do not possess this feature: every time a pair of traders drop out of the market, the supply and demand schedules shift, something which may (or may not) change the set of competitive equilibrium prices.\footnote{Remarkably, this issue was noticed in the first-ever experimental study of double auctions: see footnote 6 of \cite{smith1962experimental}. Despite conceding that `the supply and demand functions continually alter as the trading process occurs', Smith asserts that it is `the \textit{initial} [supply and demand] schedules prevailing at the opening of each trading period' that are of interest to `the theorist'. However, Smith does not give us any reason for privileging the initial demand and supply schedules over any others; and in the absence of such a reason, such privileging would appear to be entirely arbitrary.}

Our initial set of experiments yields three main findings. First, contrary to the predictions of competitive equilibrium theory, our downward shifts in the valuations do markedly decrease observed prices. Second, and partly as a result of the first finding, prices are almost never at the competitive equilibrium. Remarkably, this is the case even for our symmetric treatment, which might have been expected to yield competitive equilibrium prices. Third, prices show very little sign of converging to competitive equilibrium. Taken together, these findings imply that competitive equilibrium is not a good description of double auctions with stationary value distributions. 

We then discuss why decreasing valuations and costs might decrease observed prices. While we do not attempt to identify one model as definitively `correct', we do identify a number of models that can rationalise our finding. To take one example, within a zero intelligence model \citep{gode1993allocative}, decreasing valuations leads to stochastically lower distributions of bids and asks, thereby lowering expected prices. More interestingly, perhaps, this effect is also generated by optimising models, including those based both on myopic pay-off maximisation \citep{gjerstad1998price} and more sophisticated optimal stopping \citep{friedman1991simple}.

Our first set of experiments establishes that, if value distributions are held fixed, double auctions need not produce competitive equilibria; and that double auction outcomes are sensitive to the kind of value shifts described previously. It is natural to wonder, however, whether our shifts are still able to move observed prices in more standard double auction environments in which players are allowed to drop out of the market (without replacement) as trade progresses. To investigate this question, we also run a series of double auctions without queues, including some very long sessions involving over an hour's trading in order to give the auctions the best possible chance of equilibrating.

Our findings from these more conventional experiments are more mixed. On the one hand, there is still some evidence that our shifts depress the observed prices; and some of the sessions we run fail to equilibrate even after many periods of trading. On the other hand, there is now a marked tendency towards equilibrium, and some of our sessions do converge to equilibrium despite the marked asymmetry in the designs. Thus, these sessions reveal both the power of our shifts as well as the equilibrating forces first observed in \cite{smith1962experimental}.

Our central pair of findings --- that one can easily `break' competitive equilibrium in environments with a queue of buyers and sellers but much less easily in an environment without a queue --- suggests that the question of whether the value distribution is held fixed is of critical importance. We provide a theoretical explanation as to why this should be the case. We assume that trade follows a `Marshallian path' \citep{brewer2002behavioral}, which means that (i) at any point in time, trade takes place between the active buyer with the highest valuation and the active seller with the lowest cost, and (ii) trade occurs if and only if it is mutually beneficial (we provide a formal definition in Section \ref{The Marshallian Path}). We prove that in standard double auction formats, a Marshallian path implies that final trades must take place at equilibrium prices; and we discuss why this result might also extend to non-final trades. Importantly, this result does \textit{not} extend to double auctions with fixed value distributions, which can help explain why such auctions lack the standard equilibrating tendencies. Therefore, we identify the Marshallian path dynamic as a key driver of equilibration in standard double auctions, thus helping to solve the `scientific mystery' introduced by Vernon Smith 60 years ago.

The remainder of this article is structured as follows. Section \ref{CE preserving shifts} generalises and formalises the idea of competitive equilibrium preserving shifts. Section \ref{Experimental Design} outlines the design of experiments aimed at testing the impacts of such shifts; Section \ref{Results} contains the results of such experiments; and Section \ref{Understanding the monotonicity} discusses which models can rationalise our findings. Section \ref{The Marshallian Path} uses a combination of theory and further experimentation to study whether our findings change once value distributions are no longer held stationary. Finally, Section \ref{Concluding remarks} concludes with an outline of some new areas of research opened up by this work.

\section{CE preserving shifts}\label{CE preserving shifts}

In this section, we generalise and formalise the example from the introduction in order to obtain a better understanding of its structure. To this end, let us consider a unit mass of buyers, each indexed by $i \in [0, 1]$.\footnote{While we work with a continuum of buyers and sellers for convenience, similar results are available in the discrete case.} Each buyer has a valuation $v_i \in [0, \bar{v}_b]$ (where $\bar{v}_b > 0$ is the maximum buyer valuation) and chooses to buy (exactly) one unit of the good if and only if their valuation is at least the market price $p$. The distribution of buyer valuations is described by the cumulative distribution function $F \colon \mathbb{R} \rightarrow [0, 1]$. For simplicity, we assume that (i) $F$ has full support on the interval $[0, \bar{v}_b]$ (ii) $F$  is continuous.
 
Let $d(p)$ denote market demand at price $p$. Then
\begin{equation}
    d(p) = \int_0^1 \mathbbm{1}(v_i \geq p) di = \mathbb{P}(v_i \geq p) = 1 - F(p)
\end{equation}
where $\mathbbm{1}(v_i \geq p)$ is an indicator function. Since, $d(p) = 1 - F(p)$, $d(0) = 1 - F(0) = 1$ and $d(\bar{v}_b) = 1 - F(\bar{v}_b) = 0$. That is, demand starts at $1$ (at a price of $0$) and eventually falls to $0$ (at a price of $\bar{v}_b$). In addition, observe that $d$ is continuous (since it inherits the continuity of $F$) and strictly decreasing over the interval $[0, \bar{v}_b]$ (since $F$ has full support on this interval). In summary, then, we obtain a continuous and strictly decreasing demand function which starts at 1 before falling to 0 when the price equals the maximum valuation.

We treat sellers entirely symmetrically. That is, we have a unit mass of sellers, indexed by $i\in [0, 1]$; and each seller has a valuation (or `cost') $v_i \in [0, \bar{v}_s]$ (where $v_s > 0$ is the maximum seller valuation). Seller $i$ chooses to sell their one unit of the good if and only if $v_i \leq p$. The distribution of seller valuations is described by the cumulative distribution function $G \colon \mathbb{R} \rightarrow [0, 1]$. As before, we assume that (i) $G$ has full support on the interval $[0, \bar{v}_s]$ (ii) $G$ is continuous.

Let $s(p)$ denote market supply at price $p$. Then
\begin{equation}
    s(p) = \int_0^1 \mathbbm{1}(v_i \leq p) di = \mathbb{P}(v_i \leq p) = G(p)
\end{equation}
where $\mathbbm{1}(v_i \leq p)$ is again an indicator function. Since $s(p) = G(p)$, observe that $s(0) = G(0) = 0$ and $s(\bar{v}_s) = G(\bar{v}_s) = 1$. Moreover, $s$ is continuous (since $G$ is continuous) and strictly increasing over the interval $[0, \bar{v}_s]$ (since $G$ has full support on that interval).

We define a \textit{competitive equilibrium price} as a price $p^* \in \mathbb{R^+}$ such that $d(p^*) = s(p^*)$. According to competitive price theory, this is the price which will prevail in a market; and the associated quantity traded will be $d(p^*) = s(p^*)$. As an aside, we notice that, although competitive price theory gives us a clear prediction as to the price which should prevail in a market, it does not provide us with an explanation as to \textit{why} such a price should arise. Whether such an explanation can be provided is itself an interesting question.\footnote{\cite{aumann1964markets} and \cite{cripps2006efficiency} are particularly influential attempts to provide a foundation to competitive equilibrium theory. The result we prove in Section \ref{The Marshallian Path} provides a rather different (and much simpler) foundation, albeit one that is most directly applicable to final period trades.}

As our first observation, let us note that, under our assumptions, there is exactly one competitive equilibrium price $p^*$. To see this, define excess demand by $e(p) = d(p) -s(p)$ and note that excess demand is positive at a price of zero (specifically, $e(0) = d(0) - s(0) = 1 - 0 = 1$). Meanwhile, $e(\bar{v}_b) = d(\bar{v}_b) - s(\bar{v}_b) = 0 - G(\bar{v}_b)$, so excess demand becomes negative at a price of $v_b$. Given that excess demand inherits the continuity of $d$ and $s$, this means that there must be some $p^*$ such that $e(p^*) = 0$, i.e there exists an equilibrium price. Moreover, since $e$ is strictly decreasing, we see that $p^*$ is unique.

We now define the central concept of this section. 

\begin{definition}
A \textit{competitive equilibrium preserving demand contraction} is a transformation $T_b \colon [0, \bar{v}_b] \rightarrow \mathbb{R}^+$ such that
\begin{enumerate}
	\item $T(V_b) \leq V_b$ for all $V_b \leq p^* - \epsilon^-$
	\item $T(V_b) = V_b$ for all $V_b \in (p^* - \epsilon^-, p^* + \epsilon^+)$
	\item $T(V_b) \in [p^* + \epsilon^+, V_b]$ for all $V_b \geq p^* + \epsilon^+$
\end{enumerate}
for some $\epsilon^+,  \epsilon^- > 0$.
\end{definition}

As stated above, competitive equilibrium preserving demand contractions are downward shifts to the distribution of buyer valuations that satisfy three conditions. First, we require that low valuations  (specifically those below $p^* - \epsilon^-$) are weakly decreased. Second, we require that there is some (possibly asymmetric) $\epsilon$-ball around $p^*$ at which valuations remain unchanged. Finally, we require that high valuations (those above $p^* + \epsilon^+$) are reduced, but not reduced so much that they are brought below $p^* + \epsilon^+$. Altogether, this amounts to a stochastic reduction to the distribution of buyer valuations, although one that leaves the ranking (i.e. percentile) of valuations in a neighbourhood of $p^*$ unchanged.

We define a competitive equilibrium preserving decrease in the seller valuations analogously. 

\begin{definition}
A \textit{competitive equilibrium preserving supply expansion} is a transformation $T_s \colon [0, \bar{v}_s] \rightarrow \mathbb{R}^+$ such that
\begin{enumerate}
	\item $T(V_s) \leq V_s$ for all $V_s \leq p^* - \epsilon^-$
	\item $T(V_s) = V_s$ for all $V_s \in (p^* - \epsilon^-, p^* + \epsilon^+)$
	\item $T(V_s) \in [p^* + \epsilon^+, V_s]$ for all $V_s \geq p^* + \epsilon^+$
\end{enumerate}
for some $\epsilon^+,  \epsilon^- > 0$.
\end{definition}

\begin{figure}[H]
    \centering
    \caption{A CE preserving shift}
    \vspace{-0.3em}
    \hspace{0.2em}
    \includegraphics[width=14cm]{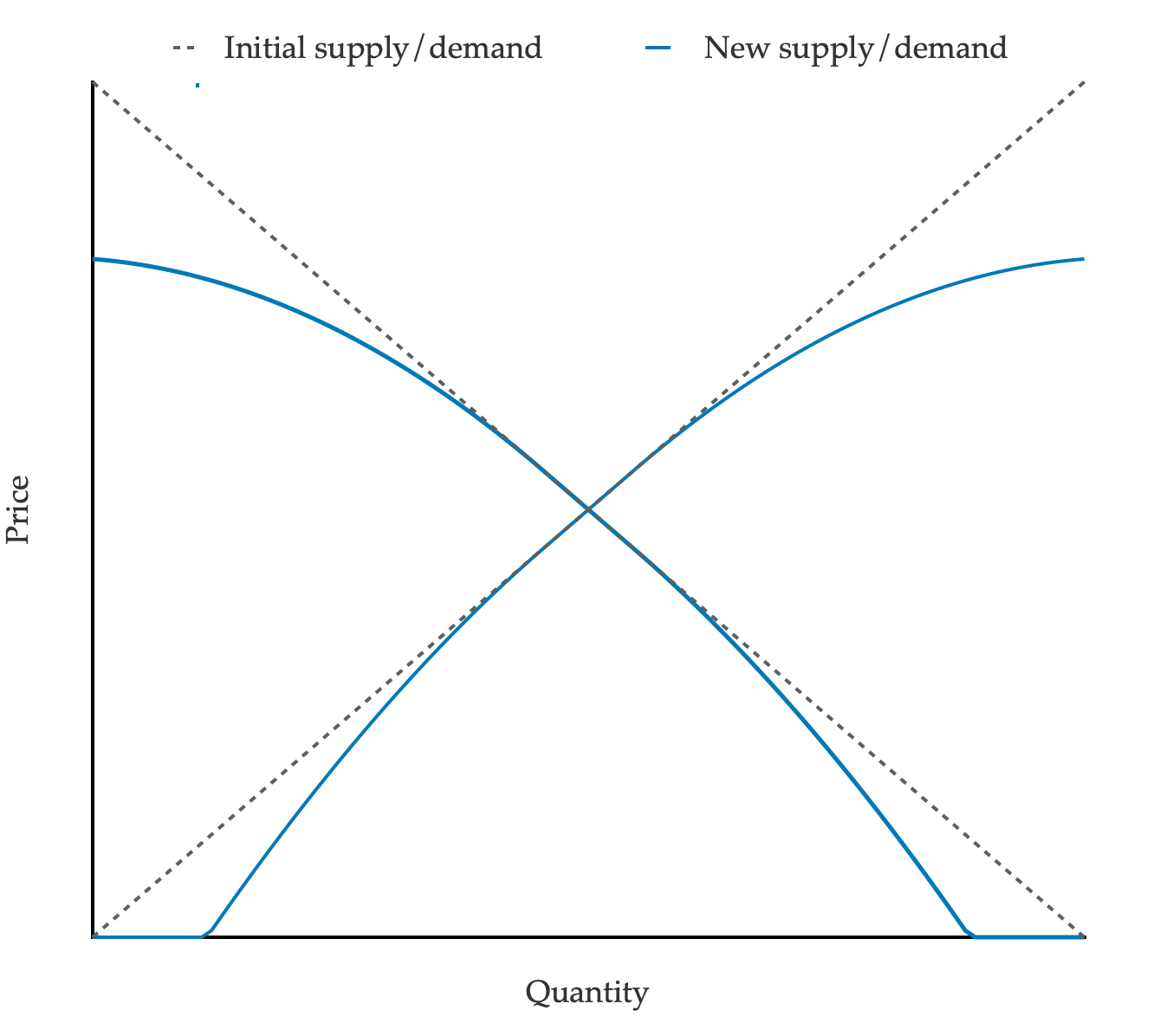}\label{shifts}
    \label{fig2}
\end{figure}

\vspace{-1em}

Figure \ref{shifts} plots a competitive equilibrium preserving demand contraction and supply expansion. As can be seen, we have decreased the buyer and seller valuations when the values were previously low. This corresponds to a downward shift in the demand and supply functions in the leftward portion of the diagram. In addition, we have left the buyer and seller valuations around the equilibrium price unchanged, which means that the demand and supply schedules remain unchanged in a neighbourhood of $p^*$. Finally, we have decreased valuations that were initially very high, corresponding to a downward shift in supply and demand in the rightward portion of Figure \ref{shifts}. As can be seen from the diagram, the unique equilibrium price remains at $p^*$ despite these downward shifts.\footnote{As will be clear from Figure \ref{shifts}, we hold valuations fixed within an $\epsilon$-ball of $p^*$ purely to preserve the uniqueness of the competitive equilibrium price. In fact, only one half of the ball is required for this purpose, although we retain the full ball for simplicity of exposition.} We now verify that this result holds in general.\footnote{All proofs are collected in Appendix \ref{proofs}.}

\begin{proposition}\label{prop1}
Let $p^*$ denote the competitive equilibrium price when buyer and seller valuations are distributed according to $V_b$ and $V_s$ respectively. Then if $T_b$ (respectively, $T_s$) is a competitive equilibrium preserving demand contraction (supply expansion), $p^*$ remains the unique competitive equilibrium price when buyer and seller valuations are distributed according to $V_b' = T_b(V_b)$ and  $V_s' = T_s(V_s)$.
\end{proposition}

We have thus seen that there is a wide class of downward shifts to the buyer and seller valuations that leave competitive equilibrium predictions unaffected.\footnote{Unsurprisingly, one can also define an analogous class of \textit{upward} shifts to the buyer and seller valuations that leave equilibria unaffected. More generally, any shift that preserves the ranking of valuations in a neighbourhood of $p^*$ unaffected will preserve the unique equilibrium price $p^*$ (although we focus on everywhere upward or everywhere downward shifts since these generate the most counterintuitive implications).} Intuitively, one might expect such shifts to depress prices, either because they encourage buyers to offer lower prices or accept lower offers from sellers, or otherwise because they encourage sellers to offer lower prices or accept lower prices from buyers. That such shifts should leave the predictions of competitive equilibrium unchanged may thus come as a surprise. Whether these counterintuitive predictions are borne out by the data is a topic that we take up in the next section.

\section{Experimental design}\label{Experimental Design}

In order to examine the effect of competitive equilibrium preserving shifts, we ran a series of double auction experiments in Oxford in early June 2022.\footnote{The experiments received IRB Approval from the University of Oxford (ECONCIA21-22-44) and were pre-registered on the AEA registry: \url{https://www.socialscienceregistry.org/trials/9547}} The basic idea of the experiment was straightforward. Buyers and sellers were first endowed with their own (private) values and costs using the technique of induced valuations (see \cite{smith1976experimental} for discussion and elaboration). They then participated in a series of double auctions. In such auctions, buyers may, at any point in time, make `bids' to purchase at a particular price or accept offers that have been made by sellers. Similarly, at any point in time, sellers may offer to sell at a particular price (`making an ask') or accept a bid made by a buyer.

We opted to conduct a series of oral double auctions, which means that subjects made bids/asks verbally instead of submitting them electronically. While this resulted in somewhat slower data collection than one would have obtained from a computerised experiment, it yielded several important advantages. First, based on some pilot experiments, it seemed that subjects found oral double auctions more engaging, and also found the structure of oral double auctions rather easier to understand. Second, using oral double auctions ensured that our results were maximally comparable to classic studies like \cite{smith1962experimental}, \cite{smith1965experimental} and so forth. For this reason, we kept as close as possible to classical experimental economics protocols (see \cite{plott_document} for a helpful document outlining how such experiments were run and \cite{kimbrough2018testing} for a more recent experiment that adheres closely to such protocols).

In this initial set of experiments, we employed a queue in order to keep supply and demand schedules stable over time. This meant that every time a trade was executed, a new buyer and seller entered the market with the valuation and cost of the just departed buyer and seller. As discussed in the introduction, we did this for two reasons. First, such a design is arguably more realistic: actual markets do not typically dissolve after several trades have occurred (as in standard double auction experiments). Instead, they are continually replenished by a steady flow of new buyers and sellers. Second and much more importantly, our queue ensured that the set of competitive equilibria remained fixed over time, allowing us to rigorously study whether prices approached \textit{the} competitive equilibrium set. Standard experiments do not necessarily have this property.

To implement the queue experimentally, we recruited a group of four buyers and four sellers for every session who began each trading period as `inactive' (so they could not engage in market activity). As trade progressed, buyers and sellers were successively drawn from the queue into the main trading area, and explicitly adopted the value and cost of the buyer and seller who had just departed (by sitting down in their place and inspecting the back of the value/cost card that had been left on the table). This was done in full view of the other experimental participants so as to emphasise that the distribution of the values and costs had remained unchanged.

To study the impact of competitive equilibrium preserving shifts, we used two different treatments, each with five `active' buyers and five `active' sellers (in addition to the eight initially inactive traders in the queue). In the `symmetric' treatment, buyer valuations were £12, £32, £52, £72, £92; and seller valuations were £8, £28, £48, £68, £88. One can check that this yields an interval of competitive equilibrium prices from £48 to £52.\footnote{The competitive equilibrium prices of £48 and £52 are only \textit{weak} competitive equilibria: at such prices, there exists at least one trader who lacks any strict incentive to act in a way that clears the market. However, prices of £49, £50 and £51 are strict equilibria.} Our second treatment (the `low values' treatment) was obtained by decreasing the valuations and costs in the symmetric treatment as aggressively as possible in a way that preserves the set of competitive equilibrium prices. To this end, we changed the buyer valuations of £12 and £32 to £0 and reduced the buyer valuations of £72 and £92 to £52. Meanwhile, we reduced the seller valuations of £8 and £28 to £0 and reduced the seller valuations of £68 and £58 to £52. This yielded the new vector of buyer and seller valuations, namely £0, £0, £52, £52, £52 for the buyers and £0, £0, £48, £52, £52 for the sellers. One can verify directly that these distributions yield the very same set of equilibrium prices, namely £48 -- £52.\footnote{Since the low values treatment yields most of the surplus to sellers at the competitive equilibrium, it is reminiscent of the extreme earnings inequality design that has been studied in the literature (see, for example, \cite{smith2000boundaries}). However, it differs in at least two important respects. First, our design has an equal number of buyers and sellers: in contrast, the classic design generates extreme earnings inequality through an imbalance in the number of buyers and sellers. Second, our design generates an interval of strict competitive equilibrium prices (namely, £49, £50 and £51). In contrast, the standard extreme earnings design leads to the non-existence of strict competitive equilibria (the weak competitive equilibrium is computed by finding the price at which one side of the market is indifferent between trading and not trading).}

To control for subject fixed effects, we conducted both treatments sequentially within every experimental session. To get a handle on order effects, we conducted two sessions (Sessions 1 and 2) and varied the order of the treatments within these sessions. We ran the symmetric treatment first in Session 1; and ran the low values treatment first in Session 2 (see Table \ref{overview} for an overview of all experimental sessions).

At the start of each session, the auction rules were presented in written form (see Appendix \ref{instructions} for the rules with which subjects were presented). The rules were then further emphasised through an extensive oral quiz; and subjects were asked if they had any outstanding questions about the auction rules. Finally, subjects were asked to engage in a mock round of trading (which would not be used to calculate payoffs) for didactic purposes. As a result of these measures, subjects' understanding of the rules seemed to be excellent. As one indication of this, 99.5\% of bids and asks made in our experiments were `individually rational' in the sense that they would have made a (non-negative) profit if accepted. This compares favourably to existing auction datasets: for example, the dataset used by \cite{lin2020evidence} involves individual rationality violations in 90\% of rounds (see \cite{ledyard_unpublished} for discussion).\footnote{In our main analysis, we drop the handful of bids and asks that violate individual rationality from our dataset. However, our results are almost entirely unaffected if we include such data points.}

Once the illustrative trading round had concluded, the real trading began. Within each round of trading, active buyers and sellers were free to make or accept offers (`bids' or `asks') at any time. All offers were repeated by the auctioneer and recorded on a whiteboard.\footnote{To make or accept an offer, a buyer would say (for example) `Buyer 2 bids 30' or `Buyer 2 accepts 60'; and this would be duly repeated by the auctioneer. An analogous comment applies to sellers. All offers and acceptances were recorded by a research assistant and double-checked using an audio recording of the experimental sessions.} Trading used the standard improvement rule, which meant that bids needed to get successively higher and asks needed to get successively lower until a transaction occurred (at which point everything was reset and all bids and asks became permissible). Each round continued until the queue had been exhausted (i.e. until four trades had occurred); and we conducted five rounds of trading for each of the two sets of value distributions. In line with recent recommendations \citep{charness2016experimental, azrieli2018incentives} and double auction experiments \citep{ikica2018competitive}, subjects were only paid for one randomly chosen round within a session.\footnote{On average, subjects received £18.36, with a mean absolute deviation of £12.09. In the experiments reported in Section \ref{The Marshallian Path}, average earnings were £16.98 with a mean absolute deviation of £11.35. In general, sessions took around 1.5 hours, about half an hour of which was devoted to carefully explaining the rules to subjects, with the remaining time devoted to trading.}

\section{Results}\label{Results}

\begin{figure}
\centering

\begin{subfigure}{Panel A (Session 1)}
\centering
\includegraphics[width=\linewidth]{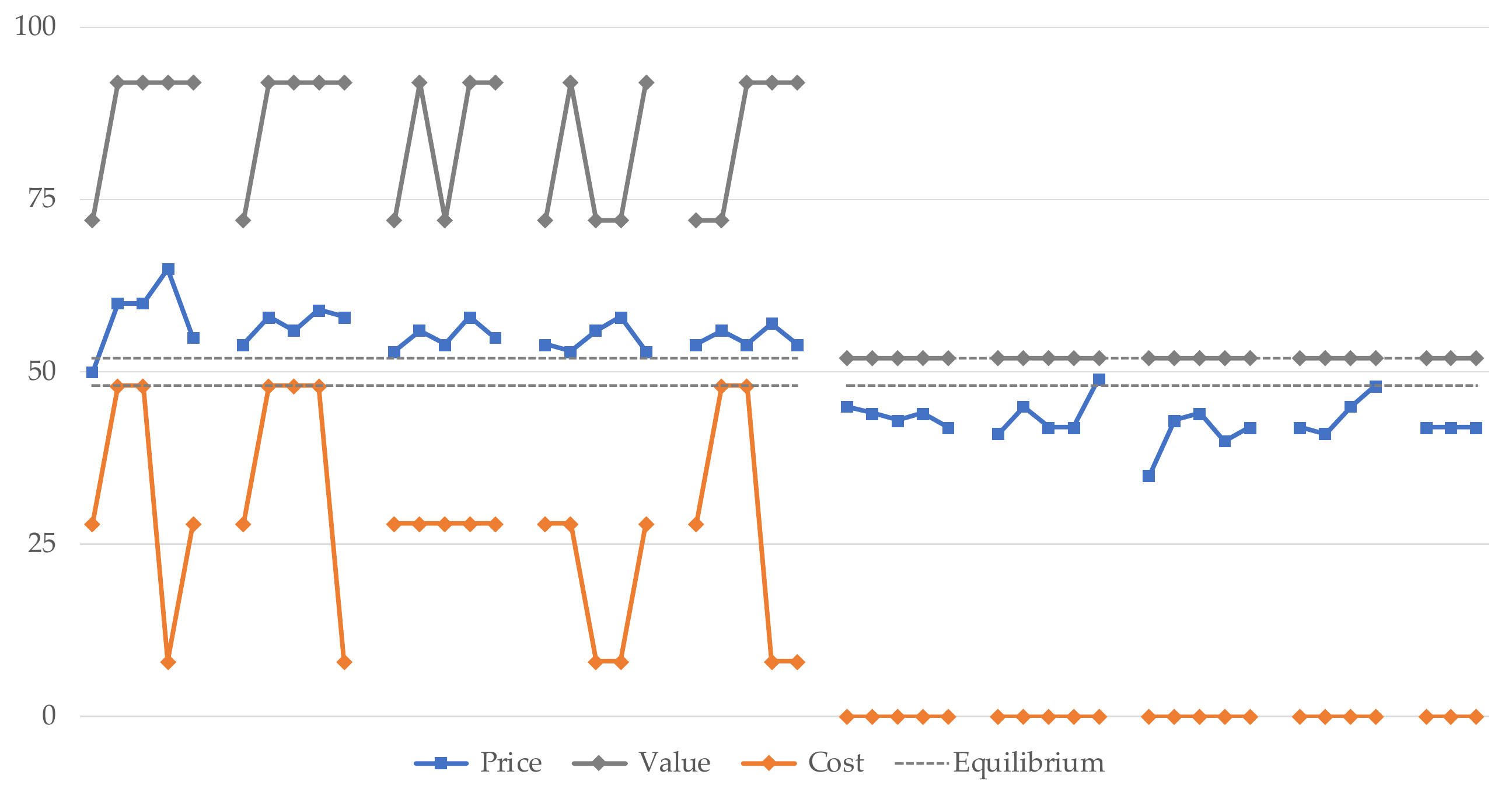}
\end{subfigure}
\begin{subfigure}{Panel B (Session 2)}
\centering
\includegraphics[width=\linewidth]{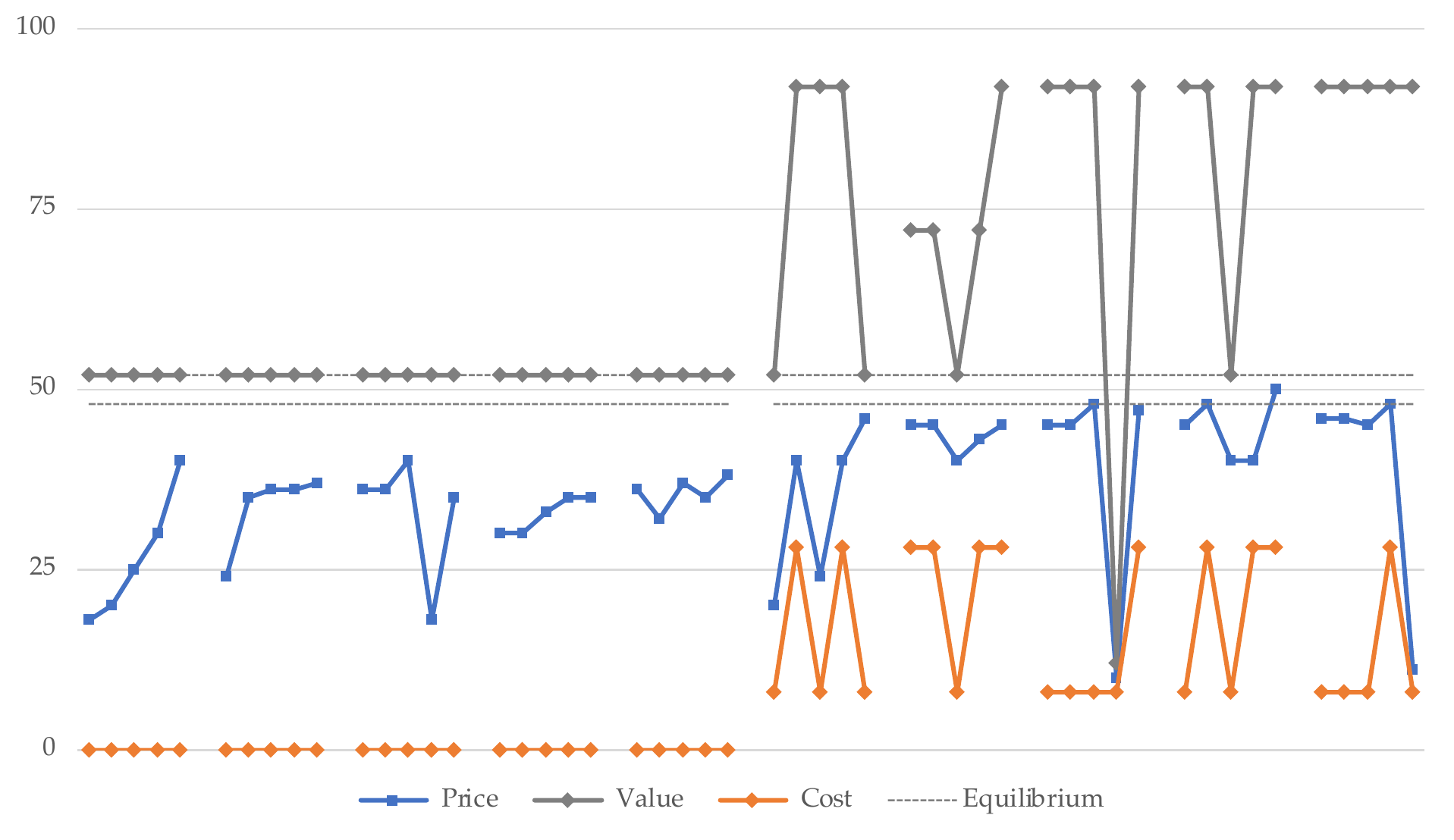}
\end{subfigure}

\medskip

\begin{minipage}[t]{0.2\textwidth}
\caption{Valuations, prices, and costs}\label{sessions_1_2}
\end{minipage}

\end{figure}

We begin by examining the transactions that occurred in our first two experimental sessions. Panel A of Figure \ref{sessions_1_2} displays the buyer valuation (top line), price (middle line) and cost (bottom line) associated with each of the transactions in Session 1. The band of equilibrium prices (£48--£52) is indicated by the dotted lines, and the end of each of the five rounds is indicated by a break. The left half of Panel A shows the transactions from the first half of the experiment (i.e., the symmetric treatment); whereas the right half of Panel A shows the transactions from the second half of the experiment (i.e. the low values treatment). Analogously, Panel B of Figure \ref{sessions_1_2} displays the valuations, prices and costs associated with the transactions from Session 2, which began with the low values treatment and proceeded to the symmetric treatment.

Three results are apparent. First, it is clear that shifting valuations and costs downward lowers observed prices, in violation of competitive equilibrium. Comparing the first halves of the separate sessions --- which is perhaps the cleanest comparison since it is uncomplicated by order effects --- we see that average prices are £56.0 (in the symmetric treatment) as opposed to £32.3 (in the low value treatment) ($p < 0.0001$).\footnote{The $p$-values in this and the next paragraph are generated by unpaired $t$-tests of the hypothesis of equal means.} We also see a similar trend within sessions. In the first session, shifting values and costs downward reduces average bids from £56.0 to £42.9 ($p < 0.0001$). In the second session, shifting values and costs upward increased average bids from £32.3 to £40.0 ($p < 0.01$). Therefore, we obtain strong evidence that competitive equilibrium preserving shifts do in fact shift prices, though the effects are substantially larger when comparing across sessions than when comparing within sessions (as one might expect given the `price stickiness' observed in the data).

While Figure \ref{sessions_1_2} solely displays data for the transactions, we can also see a similar pattern when examining the data on bids and asks. Comparing the first halves of the two sessions, we see that average bids/asks are £47.0/£67.9 in the symmetric treatment, as opposed to £26.6/£58.9 in the low values treatment ($p < 0.0001$, $p = 0.03$).\footnote{In all our analyses of the bid and ask data, we remove the handful of (rather optimistic) asks that exceed £1,000 (e.g. we drop one participant's offer to sell for £1 million). Including these asks only strengthens our conclusions.} In the first session, shifting values downward reduces average bids from £47.0 to £38.0  ($p < 0.01$) and reduces average asks from £67.9 to £48.5 ($p < 0.0001$). In the second session, shifting values up increases average bids from £26.6 to £31.7 ($p = 0.03$) and increases average asks from £59.0 to £153.2 ($p = 0.02$). We therefore conclude that, just as shifting valuations and costs downward reduces observed transaction prices, it also tends to reduce the bids and asks made by traders.

Our second main finding, which partially although not entirely follows from the first, is that prices are almost never at competitive equilibrium. In the first session (again, see Figure \ref{sessions_1_2}), prices start consistently above equilibrium, and then fall persistently below it. Moreover, not merely does competitive equilibrium fail as a literal description of the observed prices, but we can also reject a stochastic version of competitive equilibrium that allows for independent errors in every period ($p < 0.0001$, $p < 0.0001$).\footnote{To test this, we evaluate the null hypothesis that the price data were i.i.d. draws from a normal distribution with a mean of the closest competitive equilibrium price and a variance to be estimated from the data. Observe that both of these choices --- choosing the closest competitive equilibrium price along with allowing the variance to be fit ex post --- substantially stacks the deck in competitive equilibrium's favour, which then makes the clear rejection of competitive equilibrium even more striking.} The prices in our second session are also almost never at equilibrium, but in a different way. Now, prices are persistently below even the lowest competitive equilibrium price. Again, not only can we reject a rather literal interpretation of competitive equilibrium, but we can also reject a stochastic version that allows for independent errors ($p < 0.0001$, $p < 0.01$).

Our third main result is that prices do not seem to be converging to competitive equilibrium over time. In Session 1, there is little indication that prices are trending downwards (in the first half) or upwards (in the second half). Indeed, the average price changes are close to zero (£0.17 and -£0.14 in the first and second half respectively) and neither are statistically different from zero ($p = 0.84$, $p = 0.89$). In Session 2, there is a little more indication of an upward drift in prices, but again this is very weak: average changes are again close to zero (£0.83, £0.24) and again statistically insignificant ($p = 0.61$, $p =  0.94$). Even more strikingly, in neither session does competitive equilibrium seem to be an absorbing state. For example, although the price starts at equilibrium in Session 1 (with a first transaction price of £50), prices quickly drift upwards away from the equilibrium set. Similarly, although prices hit competitive equilibrium briefly in the third round of the second half of Session 2, they again move away from it. Thus, there is little appearance of convergence in about an hour's worth of trading.

In summary, we see that in double auctions with stationary value distributions (a property ensured through our use of a queue), resulting prices can remain far from competitive equilibrium even after long periods of trading, and show little sign of converging to equilibrium. Consistent with this, our competitive equilibrium preserving shifts substantially depress observed prices. In the next section, we turn to the question of what might explain this phenomenon.

\section{Understanding the monotonicity}\label{Understanding the monotonicity}

There are a number of double auction models which can rationalise the monotonicity documented in the previous section (i.e., that lower values/costs generate lower prices). To start with, consider the zero intelligence (ZI) model introduced by \cite{gode1993allocative}: buyers bid uniformly between 0 and their valuation, sellers bid uniformly between their valuation and some maximum, and trade occurs when the market bid and market ask cross (at a price equal to the earlier of the two offers). Under such assumptions, decreasing buyer and seller valuations leads to buyer bid distributions and seller ask distributions that are stochastically lower (in the sense of first-order stochastic dominance). As a result, it will tend to decrease observed prices.

To see how this works quantitatively, we conducted an extensive simulation of ZI trading under both of our experimental treatments. To operationalise ZI trading, one needs to specify a maximum ask; and we set this maximum at £100. We also assumed that at every point in time, one trader (either a buyer or seller) is chosen randomly to make an offer; and we then simulated a sequence of 10 million such offers (leading to about 800,000 market prices).\footnote{If the value distributions are held fixed as in our experiment, then nothing changes under ZI trading at the conclusion of a round. Thus, it is easiest to simply simulate a very large number of offers (and examine the resulting prices when these offers lead to trade) instead of simulating a large number of rounds.} The simulation reveals that, as one would expect, average ZI prices in the symmetric treatment are around £50. Meanwhile, average ZI prices in the low values treatment are around £35. Thus, the ZI model predicts that shifting values and costs down in the way done in our experiments should very substantially depress average prices.

While ZI can rationalise the monotonicity we observe in our data, it is doubtful that postulating random bids and asks can be said to explain the source of the monotonicity in any meaningful way. Fortunately, however, such monotonicity is also generated by optimising models. For instance, consider the model developed in \cite{gjerstad1998price}: buyers and sellers choose bids and asks in a way that maximises this period's expected pay-off. Ignoring integer constraints, the optimal bid/ask satisfies a first order condition, inspection of which reveals that optimal bids/asks are increasing in valuation/costs. Thus, the Gjerstad/Dickhaut model also predicts the monotonicity observed in our data.


Finally, we observe that this monotonicity also arises in more complicated optimising models. For example, consider the model of \cite{friedman1991simple}, in which buyers and sellers optimally choose reservation prices so as to balance the benefit of waiting for better bids/offers against the costs of running out of time. As Friedman remarks (p. 60), optimal reservation prices are monotone in valuations: this means, for example, that buyers with lower valuations are happy to accept lower offers. As a result, Friedman's optimal stopping logic also predicts the monotonicity that we observe.

In this section, our goal is not to select one model which is the `true' explanation for the observed monotonicity; and still less to discuss which model can best explain all aspects of double auction experiments (for efforts in this direction, see \cite{cason1996price} and \cite{ledyard_unpublished}). Rather, our goal is simply to argue that the observed monotonicity is nothing very mysterious: indeed, it is a simple consequence of both non-optimising as well as optimising double auction models.

\section{The Marshallian path}\label{The Marshallian Path}

Although the previous results demonstrate that double auctions need not generate equilibrium prices, one might suspect that this has something to do with the queuing procedure used in order to ensure that the value distribution remains fixed. Indeed, if traders are allowed to drop out without replacement as time progresses, then one may expect that prices approach competitive equilibrium due to a Marshallian path dynamic. While this route to equilibration has been discussed informally (see e.g. \cite{brewer2002behavioral}), we now formalise the dynamic in order to obtain a more rigorous understanding of its properties.

To this end, return to the environment in Section \ref{CE preserving shifts}, recalling that $F$ and $G$ denote the distribution of buyer and seller valuations respectively, and that $p^*$ denotes the unique competitive equilibrium price. Consider now a sequence of trades, indexed by $t \in [0, T]$. Let $v_b(t)$, $v_s(t)$ and $p(t)$ denote the buyer valuation, seller valuation, and price associated with trade $t$; and (with some abuse of notation) denote the corresponding functions by $v_b$, $v_s$ and $p$. We can now formalise the concept of a Marshallian path.

\begin{definition}
A \textit{Marshallian path} is a triple $(v_b, v_s, p)$ such that
\begin{enumerate}
	\item For all $t \in [0, T]$, $v_b(t) = F^{-1}(1-t)$ and $v_s(t) = G^{-1}(t)$.
	\item For all $i \in [0, 1]$, $i \in [0, T]$ if and only if $F^{-1}(1-i) \geq G^{-1}(i)$.
	\item For all $t \in [0, T]$, $v_s(t) \leq p(t) \leq v_b(t)$.
\end{enumerate}
\end{definition}

As stated above, a Marshallian path is a sequence of trades that satisfies three conditions. To understand the first condition, start with the simpler equation $v_s(t) = G^{-1}(t)$ and invert it to get $ t = G(v_s(t))$. This says that the fraction of values that are below $v_s(t)$ is $t$; so at time $t$, the valuation of the seller about to engage in trade is at the $t$-th percentile of the seller value distribution.\footnote{Our usage of the word `percentile’ differs from the normal usage by a factor of 100: for example, we say `0.2-th percentile’ to mean the 20th percentile.} Likewise, the other equation reads $v_b(t) = F^{-1}(1-t)$, which may be inverted to yield $t = 1 - F[v_b(t)]$. This says that the fraction of buyer valuations that are \textit{above} $v_b(t)$ is $t$; so at time $t$, the valuation of the buyer about to engage in trade is at the $(1-t)$-th percentile of the buyer value distribution. Taken together, these assumptions say that trade takes place `in order’: at every point in time, trade occurs between the buyer with the highest valuation and the seller with the lowest valuation out of those `active' traders who remain in the market.

The second condition says that for any possible unit, that unit is traded if and only if the buyer's valuation for that unit exceeds the seller's valuation for that unit (assuming that these units are traded `in order’, in line with condition 1). In other words, trades take place if and only if there is some price at which they would be mutually beneficial. We will discuss the plausibility of this claim when we analyse the experimental results of this section. 

The final condition says that, in actual fact, trade is mutually beneficial: the price of every trade lies between the valuation of the relevant buyer and the valuation (i.e. `cost') of the relevant seller. Observe that, although there is just one path of trades satisfying conditions 1 and 2, there are multiple mutually beneficial price paths. Thus, as far as prices are concerned, the Marshallian path is indeterminate; which is why we may, for reasons of pedantry, choose to speak of `a' rather than `the' Marshallian path.\footnote{On the other hand, there is a single path of \textit{trades} which counts as Marshallian, which is presumably why one finds discussion of ‘the’ Marshallian path in prior literature \citep{brewer2002behavioral, plott2013marshall}.}

\begin{proposition}\label{prop2}
If $(v_b, v_s, p)$ is a Marshallian path, then $p(T) = p^*$.
\end{proposition}

\begin{figure}[H]
    \centering
    \caption{A Marshallian path}
    \vspace{-0.3em}
    \hspace{0.2em}
    \includegraphics[width=14cm]{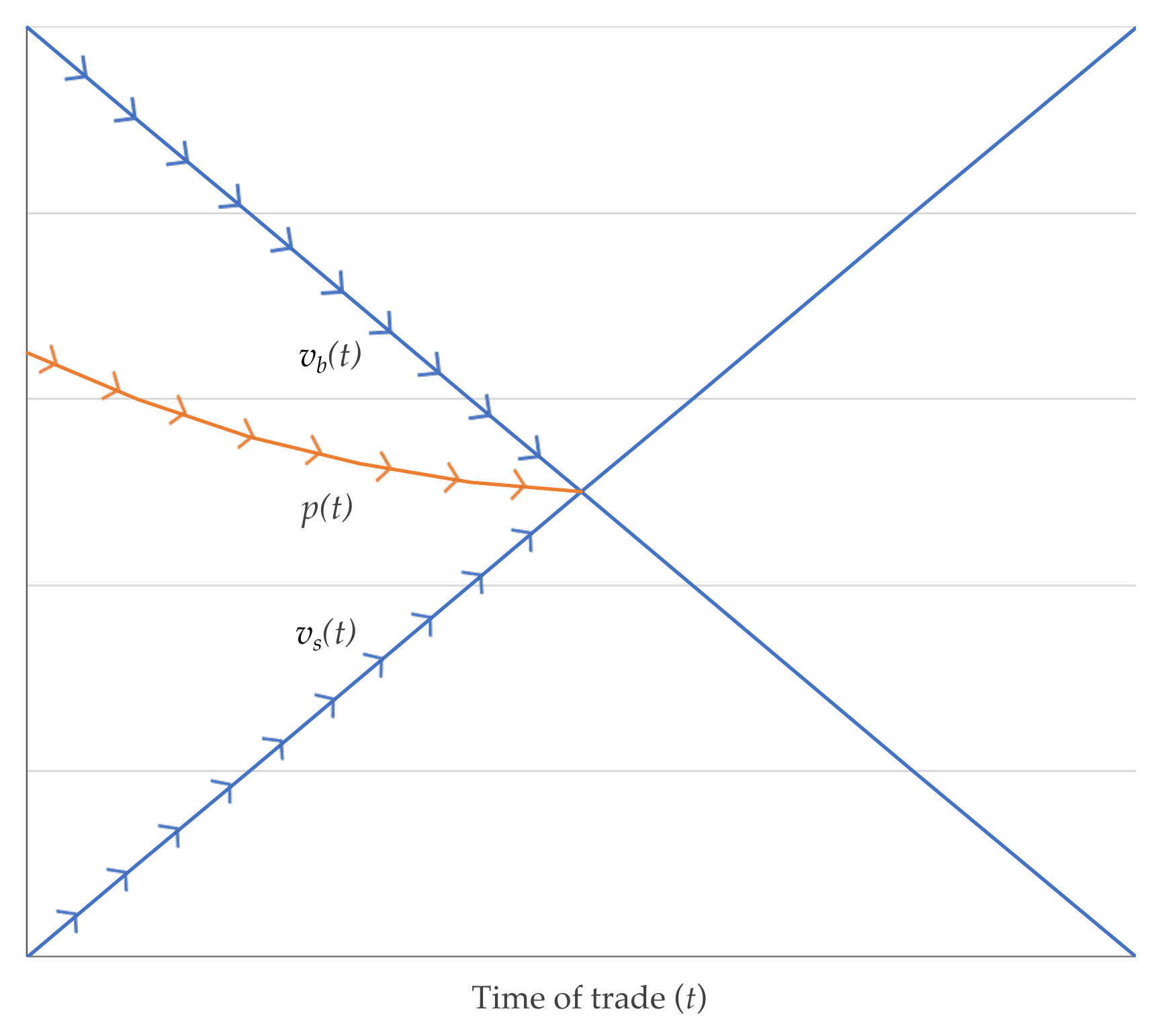}\label{fig:prop2}
\end{figure}

Proposition \ref{prop2} says that, if trade follows a Marshallian path, then the final trade must be transacted at the equilibrium price. To see why this is true, consider Figure \ref{fig:prop2}. Recall that the demand curve can be interpreted as the valuation of the marginal buyer (once buyers have been sorted from highest valuation to lowest), and the supply curve can be interpreted as the valuation of the marginal seller (assuming sellers have been sorted in order from lowest to highest valuation). As a result, to assume a Marshallian path is simply to assume that trade takes place in the rightward direction, starting with the buyer with the highest valuation and the seller with the lowest cost (condition 1). At every point in time, the price must lie between the buyer and seller valuation (condition 3), which means that the price path must lie between the demand and supply functions. Furthermore, trade occurs if and only if it could be mutually beneficial (condition 2), which means that the final trade is at the intersection of the demand and supply curves. At such a trade, the only mutually beneficial price is $p(T) =p^*$, so we obtain final period equilibration.\footnote{Incidentally, it is unclear whether this argument for equilibration actually appears anywhere in the work of Marshall. \cite{marshall1890principles} does present a theory of equilibration (see Book V, Chapter 3, Section 6), but this theory involves a quantity adjustment at disequilibrium prices. At the risk of giving Marshall credit for yet another idea that he did not originate (see \cite{humphrey1996marshallian}), we retain the phrase `Marshallian path' in deference to previous literature on this issue \citep{brewer2002behavioral, plott2013marshall}.}

\vspace{-0.7em}


The previous result establishes that if trade follows a Marshallian path, then final period prices must be at competitive equilibrium. However, it does not address why trade should take place in exactly the order required for a Marshallian path (condition 1). While we will not address this issue in detail, there are two reasons why it might seem reasonable to make this assumption. First, prior literature shows that at least as far as experimental double auctions go, condition 1 is indeed a rough approximation of reality (for instance, see \cite{plott2013marshall}). Indeed, as we later show, this assumption matches up very well with our own experimental data. Second, condition 1 should arise for precisely the reasons discussed in Section \ref{Understanding the monotonicity}: it is, after all, simply an assertion that offers are monotone in valuations. 

We should perhaps also emphasise that this result does \textit{not} extend to settings with stationary value distributions (like those studied experimentally in Section \ref{Results}). In such settings, Marshallian path logic would suggest that trade should take place between the buyer with the highest value (here, £52) and the seller with the lowest cost (here, £0): and indeed this is precisely what we observe in the low values treatment.\footnote{This phenomenon is also observed, although a little less strikingly, in the symmetric treatments.} However, as far as the logic of Proposition \ref{prop2} goes, there is no reason why they should trade at equilibrium prices if they are continually replaced once they leave the market. Moreover, under anything like an equal division of the surplus, we would expect prices to be considerably below equilibrium (as documented in Section \ref{Results}). Thus, not only does this result provide us with a reason to expect final period equilibration, but it provides us with a reason that is conspicuously absent in the case of stationary value distributions.

Motivated by our result, we now report the results of a series of additional experimental sessions held in order to determine whether the Marshallian path is indeed crucial for equilibrium convergence. These experiments followed exactly the same procedure as those outlined in Section \ref{Experimental Design}, with the exception that we removed the queue of buyers and sellers (so traders could drop out without replacement as time progressed). As in classical double auction experiments, each round now concluded once there were no longer any active traders who wanted to make or accept an offer.\footnote{Similarly to \cite{kimbrough2018testing}, the auctioneer asked: `Sellers, would any of you like to make any offers or accept the current market bid? Buyers, would any of you like to make any bids or accept the current market ask? Going once… going twice… the round has concluded'. See Appendix \ref{instructions} for the full instructions for the experiment.}

\begin{figure}
\centering

\begin{subfigure}{Panel A (Session 3)}
\centering
\includegraphics[width=\linewidth]{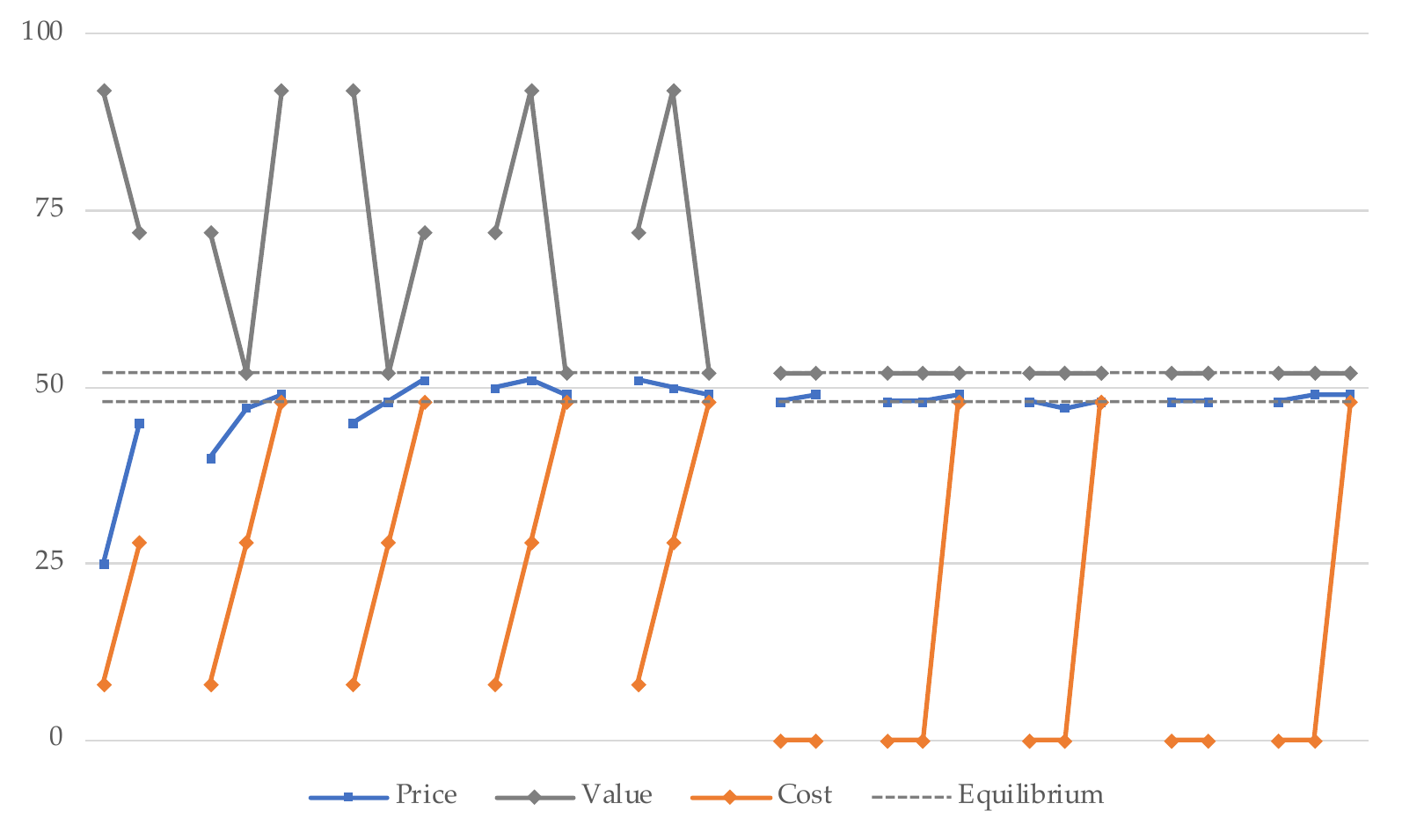}
\end{subfigure}
\begin{subfigure}{Panel B (Session 4)}
\centering
\includegraphics[width=\linewidth]{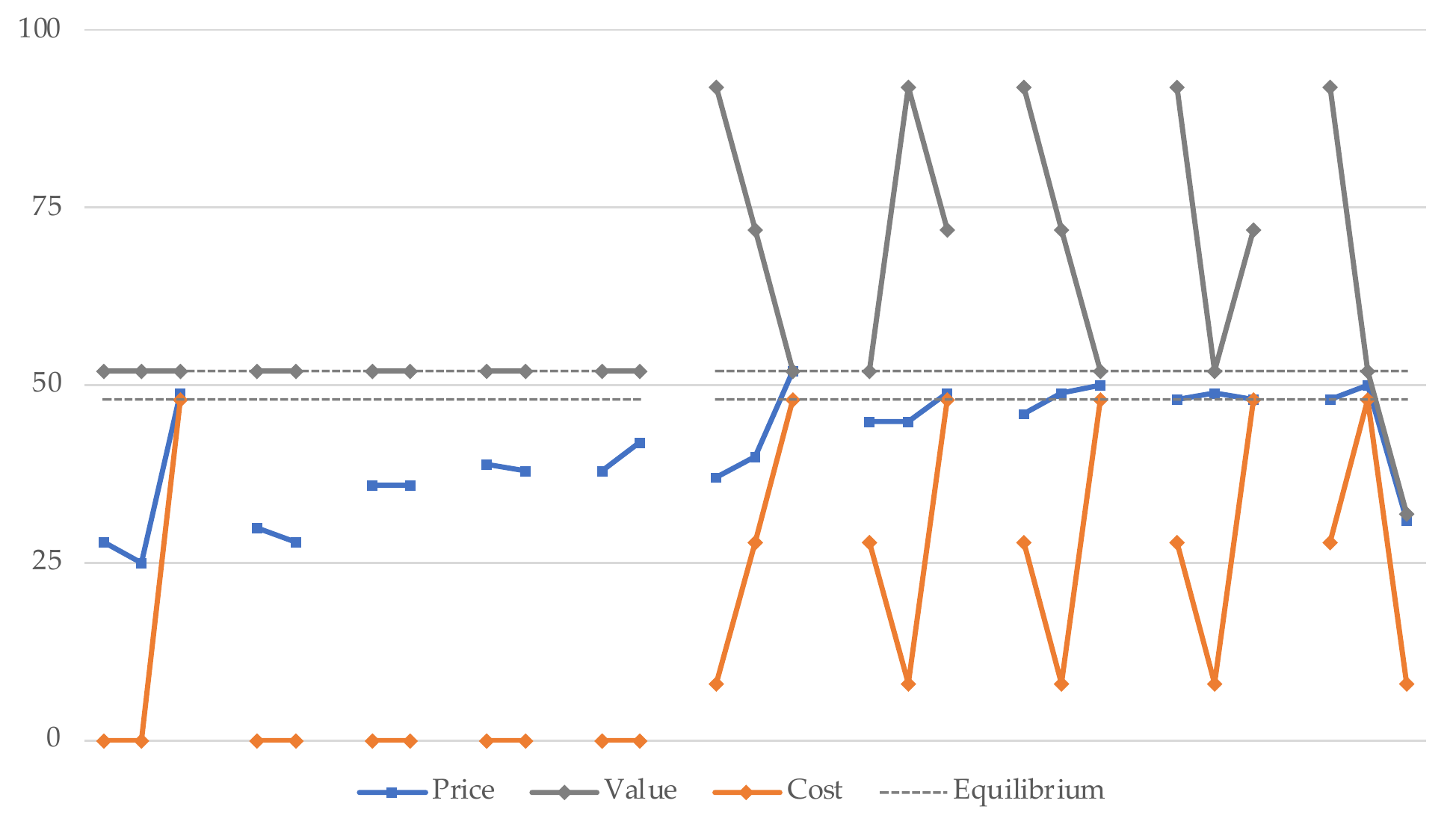}
\end{subfigure}

\medskip

\begin{minipage}[t]{0.2\textwidth}
\caption{Valuations, prices, and costs}\label{sessions_3_4}
\end{minipage}

\end{figure}

As before,  we ran two sessions (Sessions 3 and 4), varying the order in which the sessions were presented. Session 3 began with the symmetric treatment, whereas Session 4 began with the low values treatment. Panels A and B of Figure \ref{sessions_3_4} display the main results for Session 3 and Session 4 respectively. As usual, the buyer valuation, price, and seller cost associated with each transaction are denoted by the top, middle, and bottom lines respectively. The conclusion of a round is indicated by a break; and the set of equilibrium prices is indicated by the dotted lines.

Two features of the panels are especially noteworthy. First, we still obtain evidence that our shifts depress prices, although the evidence is now more mixed. Comparing across the first halves of each session, which is again the cleanest comparison since it is uncomplicated by order effects, we see that average bids are £46.4 in the symmetric treatment as opposed to £35.3 in the low values treatment ($p < 0.001)$. Thus, shifting the values downward appears to substantially depress prices, at least in the short run. Looking within periods, we see that shifting the values and costs upward increased the average bids in Session 4, from £35.4 to £45.8 ($p < 0.001$). On the other hand, no such effect is observed within Session 3: average prices are actually a little higher in the low value treatment, although the difference is not statistically significant ($p = 0.36$).

Our second finding, which again is related to the first, is that the double auctions now demonstrate strong equilibrating tendencies. In Session 3, prices reach equilibrium by the last trade of the second round of the symmetric treatment and remain within the band of equilibrium prices for the rest of the session (even after valuations are shifted downward). In Session 4, prices failed to equilibrate despite half an hour's trading in the low values treatment. However, they do show a clear upward drift, and we do obtain an equilibrium price for one transaction. Moreover, switching to the symmetric treatment leads prices to more or less equilibrate by the end of the second round.

Given the previous findings, it is natural to conjecture that the low values treatment might equilibrate (even without the help of the symmetric treatment) if it were simply given enough time to do so. To check this conjecture, we ran two further experimental sessions (Sessions 5 and 6) which solely studied the low value treatment.\footnote{These experimental sessions were not contained in our original pre-registration.} To give our auctions the best chance of equilibration, we conducted nine rounds within each session (which corresponded to around one hour of trading time per session). To prevent subjects with unfavourable valuations from becoming bored by their lack of profitable trading opportunities, we re-shuffled the values and costs twice (once at the beginning of the fourth round, and once at the beginning of the seventh). We did not explicitly inform subjects that the distribution of values and costs had remained the same, though we also gave them no indication that it had changed.

\begin{figure}
\centering

\begin{subfigure}{Panel A (Session 5)}
\centering
\includegraphics[width=\linewidth]{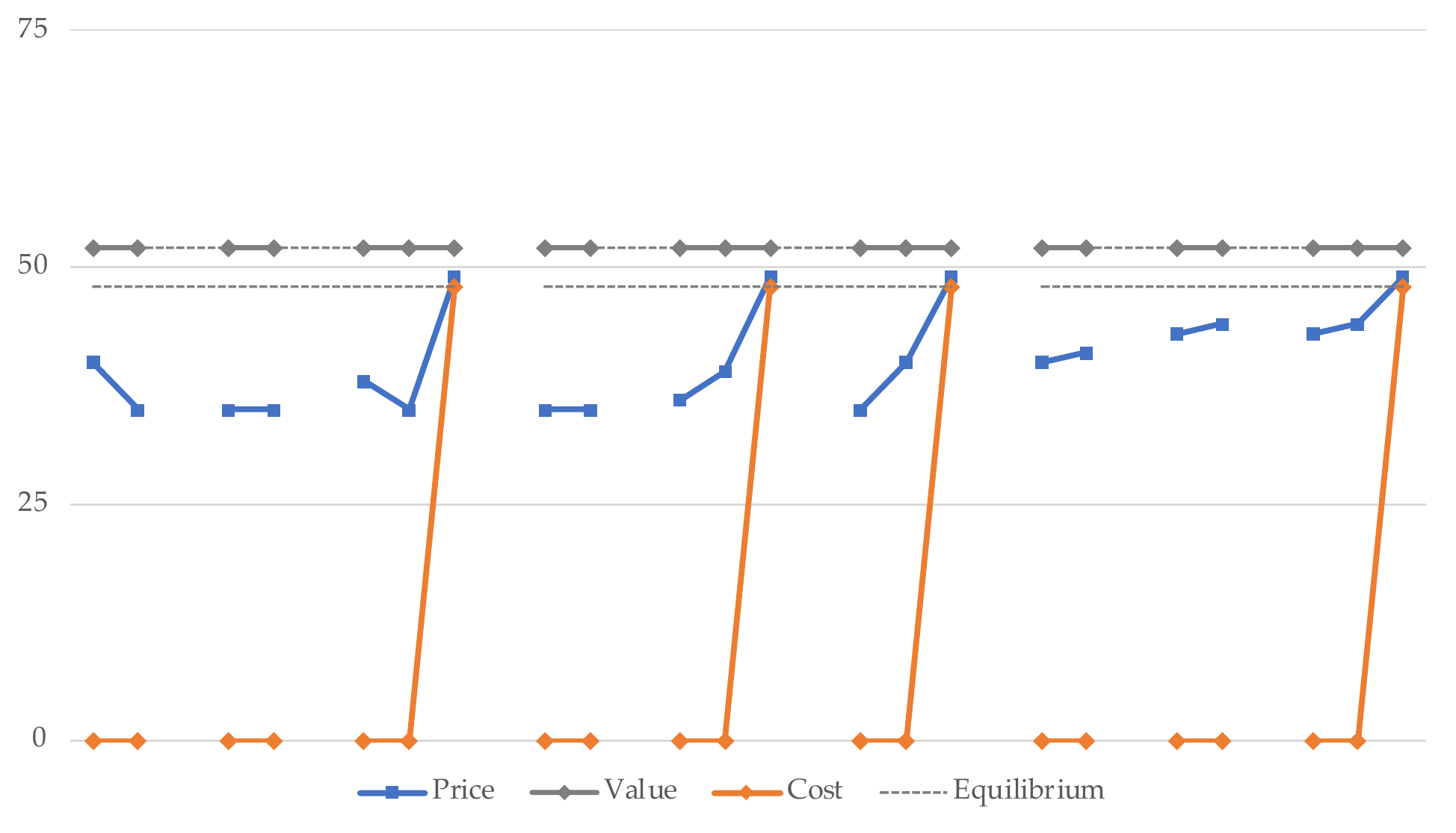}
\end{subfigure}
\begin{subfigure}{Panel B (Session 6)}
\centering
\includegraphics[width=\linewidth]{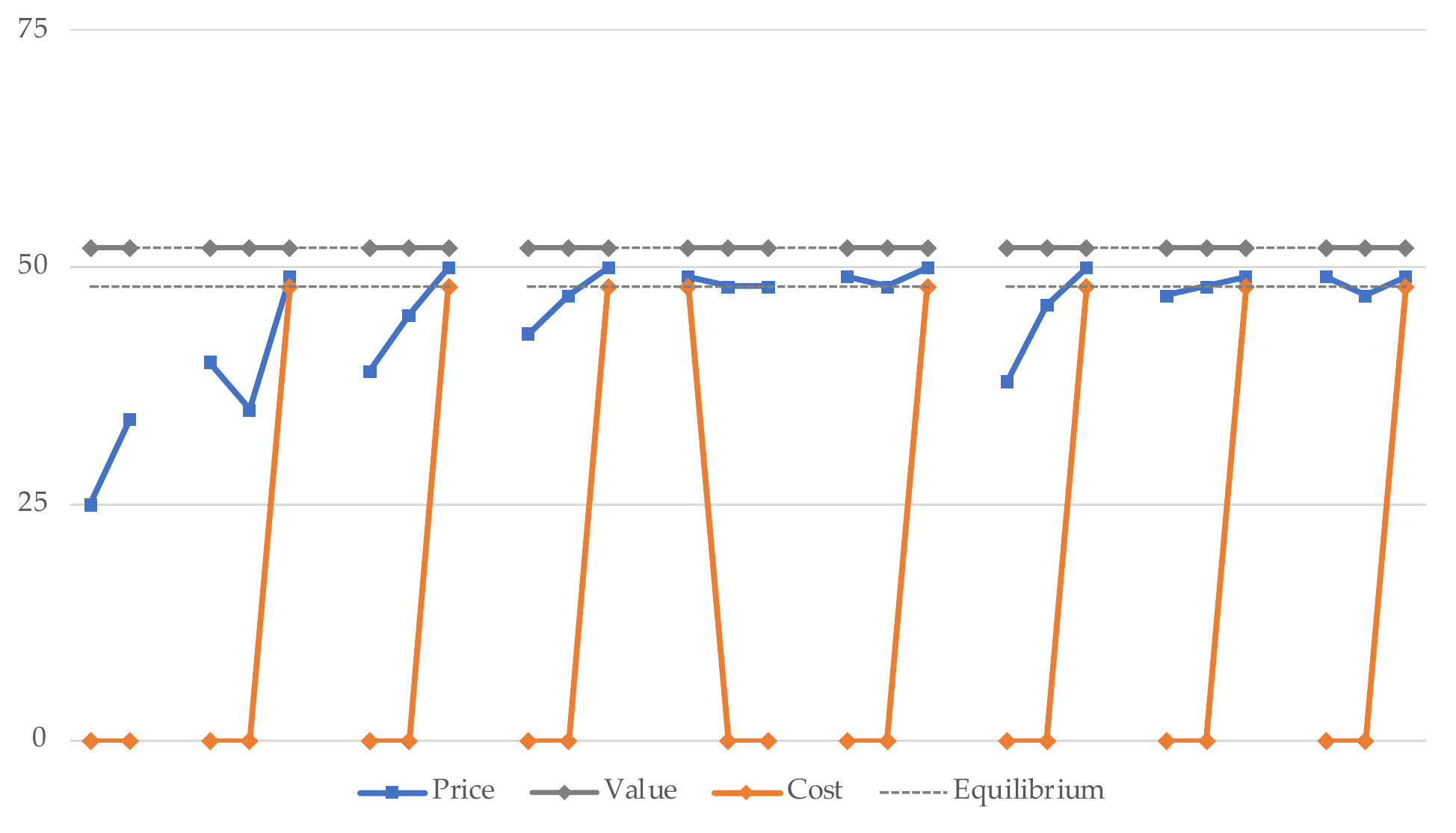}
\end{subfigure}

\medskip

\begin{minipage}[t]{0.2\textwidth}
\caption{Valuations, prices, and costs}\label{sessions_5_6}
\end{minipage}

\end{figure}

Panels A and B of Figure \ref{sessions_5_6} display the results of our final two sessions. Several features of the data are evident. First, trade very clearly follows the order assumed by the Marshallian path. We find that trade follows the exact Marshallian order in 17 out of the 18 rounds (the corresponding figure is 47 out of 48 rounds if one includes the low valuation treatments from Sessions 3 and 4). These results are consistent with previous evidence on the order of trade within double auctions (see, e.g., \cite{cason1996price}, \cite{plott2013marshall}, \cite{lin2020evidence} and \cite{sherstyuk2021randomized}).\footnote{While other studies observe a similar phenomenon in terms of trading order, their results on trading order are rather less pronounced. Presumably, this is due to the fact that the two highest valuations are the same, and the two lowest costs are also the same, making it much easier to obtain a Marshallian trading order.}

Second, we observe striking evidence of third period equilibration. As predicted by our theory, \textit{all} third trades occur at a competitive equilibrium price. Moreover, whilst not all rounds lead to a third trade, a good portion of them do. As a result, we obtain final period equilibration in 12 out of 18 of the rounds (this corresponding figure is 16 out of 38 rounds if one includes the low values treatments from Sessions 3 and 4). Consistent with our theory, prices are much less likely to be at equilibrium for non-third period trades. For example, in Session 5, we \textit{never} observe equilibrium prices for first and second period trades.

Third, we also obtain evidence that equilibration is possible even for non-final periods. This is certainly not evident in Session 5, where (as noted) we obtain equilibrium prices only for third period trades. On the other hand, in Session 6, we obtain equilibrium prices for all trades in the fifth, sixth, eighth and ninth rounds. We discuss why this may be below.

Altogether, our results are highly consistent with the theoretical result presented above. In essentially all periods, the first trades happen between buyers with the highest valuations (here £52) and sellers with the lowest cost (here £0): this validates condition 1 of the Marshallian path. Once these two trades have been executed, the only remaining mutually beneficial traders are a buyer with a valuation of £52 and a seller with a cost of £48. Thus, if all mutually beneficial trades are going to occur, a deal must be struck in the equilibrium price range of £48 -- £52. This is exactly what we see in our data (again, see Figure \ref{sessions_5_6}): it is as if a force, namely a large jump in the valuation of the marginal seller, is `squeezing' the realised price into the equilibrium set.

While our results are broadly consistent with our theory, they do leave at least two unanswered questions. The first question is why not all mutually beneficial trades are made within certain rounds. From a mechanical point of view, one can see that the obstacle is essentially on the buyers' side. In all but one of the third period trades, seller offers fall to the equilibrium range, and it is the market bid which potentially fails to rise high enough for a deal to be struck (see Figure \ref{bids_asks} for the evolution of the market bids and asks in such situations). The question is then why third period buyers with a valuation of £52 refuse to trade at equilibrium prices, thereby `leaving money on the table'. 

While we do not attempt to definitively answer this question, at least two explanations are suggested by comments made by buyers in post-experimental discussion. First, buyers may have had a strategic motive: by refusing to pay equilibrium prices, they may have been attempting to signal a reluctance to pay such prices in future rounds, thereby encouraging lower future offers from sellers.\footnote{Ironically, this would be a reason why repetition of rounds might actually hinder equilibration: such an effect would not be present in a single round double auction.} Second, buyers may have refused to accept sellers' offers of equilibrium prices on the mistaken belief that such offers would deliver almost all the surplus to the relevant seller. In rounds that generated only two trades, prices tended to be in the £35 -- £40 range, with an average of about £37. As a result, buyers might have reasonably formed the expectation that sellers' values are rather low (about £22 under the assumption of equal surplus sharing) and therefore become quite frustrated when the remaining seller started to demand competitive equilibrium prices. In such a situation, a buyer might be willing to forfeit a few pounds in profit in return for the pleasure of punishing what they take to be greediness on a seller's part (see \cite{rabin1993incorporating} and \cite{fehr1999theory} for formal models in this vein).

A final question is why prices can potentially become `stuck' at equilibrium, even for non-third period trades. For concreteness, let us consider Session 6 (see panel B of Figure \ref{sessions_5_6}) since this phenomenon does not arise in Session 5. By Marshallian path logic, one would expect all third period trades to be at equilibrium, and indeed this is precisely what we observe. However, there is no special reason, at least so far as the logic of Proposition \ref{prop2} goes, to expect equilibrium prices for earlier trades; and under anything like equal surplus sharing, one would expect the realised prices to be substantially lower. Strikingly, however, we see that once prices have been pushed to equilibrium repeatedly by the force of the Marshallian path, they may persist at equilibrium (or may not, as in Session 5). The question is then what might explain this.

While we do not attempt to answer this question formally, it is not too hard to see how the Marshallian path logic might extend to trades that do not occur in the final period. Suppose that the first two trades occur at reasonably low prices, say £35, and the third period price jumps (by the logic of Proposition \ref{prop2}) to a competitive equilibrium price. Over time, traders should notice this pattern --- and indeed many traders reported noticing this in post-experimental discussions. Given that they expect the price to rise in the final period, buyers should be especially keen to trade earlier than their competitors. This in turn should induce them to submit higher bids and to accept higher asks for early period offers. Similarly, given that the price is expected to jump in the third period, sellers should be keen to trade late, which should induce them to only offer very high asks or accept very high bids in the early periods. Taken together, these forces explain how the final period equilibration delivered by the Marshallian path might eventually drag earlier round prices up to equilibrium levels.\footnote{Importantly, a similar argument can apply even in rounds which lack a third period trade. Such third periods typically involve a substantial increase in buyer bids and seller asks relative to previous periods. As a result, even those periods that do not lead to a trade can facilitate equilibration.}

\section{Concluding remarks}\label{Concluding remarks}

In this paper, we revisit the link between competitive equilibrium and the double auction. We begin by showing that competitive equilibrium generates a set of highly counterintuitive predictions. Specifically, it predicts that prices can remain unchanged following a certain class of downward shifts to buyer and seller valuations. We find that in double auctions with stationary value distributions, these predictions are strongly falsified; and more generally that competitive equilibrium is a poor description of outcomes in such auctions. On the other hand, we also find that the effectiveness of our counterexamples is blunted in double auctions in which traders may drop out (without replacement) as time progresses. Taken together, this pair of findings imply that whether value distributions are held stationary is a crucial determinant of whether prices converge to competitive equilibrium; which in turn suggests that the Marshallian path is a key driver of equilibration in double auctions.

Despite the long history of double auction research, we believe that our findings open up several new avenues for investigation. First, further experimental work on the potential importance of the stationarity of the value distribution is clearly needed. Indeed, we are aware of only one other paper on this issue \citep{brewer2002behavioral}; and that paper reaches the rather contrasting conclusion that double auctions can converge even with stationary value distributions.\footnote{In part, the apparent difference between this paper and ours may be somewhat illusory. Examining Figures 7 -- 9 in \cite{brewer2002behavioral}, one sees that prices are persistently above equilibrium, and then (following the shifts) persistently below equilibrium. While we would not classify this as `convergence to equilibrium', it does count as convergence under the relatively undemanding definition used by  \cite{brewer2002behavioral} (see p. 191 for details).} Given the paucity of studies on this issue and the apparently mixed nature of the evidence, it is vital that further experiments are conducted to verify whether the issue of stationarity is indeed as critical as we claim. 

On the theoretical front, it may also be useful to explicitly develop models of double auction behaviour in stationary environments. Classical double auction models, e.g. \cite{easley1986theories}, \cite{friedman1991simple} and \cite{gjerstad1998price}, explicitly address the situation in which traders drop out without replacement as time progresses, and it would be interesting to investigate the extent to which their insights carry over to the stationary setting. Indeed, the stationary setting would appear to be rather more tractable from a modelling perspective, since it may be viewed (at least very roughly) as a repeated version of the much simpler $k$-double auction. As a result, the insights of \cite{chatterjee1983bargaining}, \cite{jackson2005existence}, \cite{reny2006toward}, \cite{fudenberg2007existence} and others may prove highly relevant.

Finally, it might be valuable to further study what precisely hinders equilibration in the `low values treatment' investigated in this paper. We have informally sketched two alternatives: that buyers reject equilibrium offers on the (mistaken) belief that they allocate almost all the surplus to sellers, or otherwise that buyers reject these offers strategically as a means of generating higher offers in subsequent rounds. While we have not attempted to learn which of these is the key driver, this would seem straightforward to check experimentally. For example, the second channel is possible only when rounds are repeated, so can be easily ‘turned off’ by conducting double auctions with a single incentivised round.

\clearpage

\setlength{\bibhang}{0pt}
\bibliographystyle{apalike}
\bibliography{bibliography.bib}

\newpage

\setcounter{table}{0}
\renewcommand{\thetable}{A\arabic{table}}

\setcounter{figure}{0}
\renewcommand{\thefigure}{A\arabic{figure}}

\begin{appendices}

\section{Proofs} \label{proofs}

\begin{proof}[Proof of Proposition \ref{prop1}]
We begin by arguing that $p^*$ remains \textit{an} equilibrium price when valuations are distributed according to $T_b(V_b) = V_b'$ and $T_s(V_s) = V_s'$. To this end, define $\mathcal{B} = (p^* - \epsilon^-, p^* + \epsilon^+)$ and fix some $p \in \mathcal{B}$. By the law of total probability,
\begin{equation}\label{totalprob}
\begin{split}
P(V'_b \leq p) &=  P(V'_b \leq p|V_b \leq p^* - \epsilon^-)P(V_b \leq p^* - \epsilon^-) + P(V'_b \leq p|V_b \in \mathcal{B} )P(V_b \in \mathcal{B} ) \\ &+ P(V'_b \leq p |V_b \geq p^* + \epsilon^+)P(V_b \geq p^* + \epsilon^+)
\end{split}
\end{equation}
Since $T_b(V_b) = V_b'$, where $T_b$ is a CE preserving demand contraction,
\begin{itemize}
    \item If $V_b \leq p^* - \epsilon^-$, then $V'_b \leq V_b \leq p^* - \epsilon^- \leq p$, so $P(V'_b \leq p|V_b \leq p^* - \epsilon^-) = 1$
    \item If $V_b \in \mathcal{B}$, then $V_b' = V_b$ and so $P(V'_b \leq p|V_b \in \mathcal{B} ) = P(V_b \leq p|V_b \in \mathcal{B} )$
    \item If $V_b \geq p^* + \epsilon^+$, then $V_b' \geq p^* + \epsilon^+ > p$, so $P(V'_b \leq p |V_b \geq p^* + \epsilon^+) = 0$
\end{itemize}
Inserting these equalities into (\ref{totalprob}), we obtain
\begin{equation}
\begin{split}
P(V'_b \leq p) &= P(V_b \leq p^* - \epsilon^-) + P(V_b \leq p|V_b \in \mathcal{B} )P(V_b \in \mathcal{B} ) \\
&= P(V_b \leq p^* - \epsilon^-) + P(V_b \leq p \wedge V_b \in \mathcal{B} ) \\
&= P(V_b \leq p^* - \epsilon^-) +P(p^* - \epsilon^- < V_b \leq p) \\
&= P(V_b \leq p)
\end{split}
\end{equation}
Let $F'$ and $G'$ denote the distributions of $V_b'$ and $V_s'$; let $d'(p)$ and $s'(p)$ denote the demand and supply functions generated by these distributions; and define $e'(p) \equiv d'(p) - s'(p)$.

The argument just given reveals that $F'(p) = F(p)$ for any $p \in \mathcal{B}$. Therefore, $d'(p) = 1 - F'(p) = 1 - F(p) = d(p)$ (for $p \in \mathcal{B}$). In particular, $d'(p^*) = d(p^*)$. Likewise, if we replace $F$ with $G$ in the previous argument, we see that $G'(p) = G(p)$ for any $p \in \mathcal{B}$. Hence, $s'(p) = G'(p) = G(p) = s(p)$ at such prices; and in particular, $s'(p^*) = s(p^*)$. Since $d(p^*) = s(p^*)$ by definition, this means that $d'(p^*) = s'(p^*)$, i.e. $p^*$ remains a CE.

To see that the CE remains unique, fix some $p < p^*$ and consider the following cases:

Case 1: $p \in \mathcal{B}$. Then $e'(p) = e(p)$ (shown above), and $e(p) > e(p^*) = 0$ (since $p < p^*$ and $e$ is strictly decreasing). This means that $e'(p) > 0$, i.e. no such prices clear the market.

Case 2: $p \notin \mathcal{B}$. Since $p < p^*$, this means that $p \leq p^* - \epsilon^-$. Now define $\bar{p} \equiv p^* - 0.5\epsilon^-$. Plainly, $\bar{p} \in \mathcal{B}$ and $p < p^*$, so $e'(\bar{p}) = e(\bar{p}) > 0$. Moreover, since $e'(p) = 1 - F'(p) - G'(p)$ where $F'$ and $G'$ are CDFs, $e'$ is weakly decreasing over $\mathbb{R}^+$. Thus, if $p \leq p^* - \epsilon^- < \bar{p}$, then $e'(p) \geq e'(\bar{p}) = e(\bar{p}) > 0$, so no such prices can clear the market either.

Since these cases exhaust the possibilities, we see that no prices $p < p^*$ can be CE. By an analogous argument, one can also rule out prices $p > p^*$. We conclude that $p^*$ must remain the unique competitive equilibrium price.\end{proof}

\begin{proof}[Proof of Proposition \ref{prop2}]
First, we show that $v_b(T) = v_s(T)$. To see this, recall condition 2 of the Marshallian path definition: for every $i \in [0, 1]$, $i \in [0, T]$ if and only if $F^{-1}(1-i) \geq G^{-1}(i)$. Define $\phi(i) = F^{-1}(1-i) - G^{-1}(i)$, so our inequality is $\phi(i) \geq 0$. Observe that $\phi(0) = F^{-1}(1) - G^{-1}(0) = \bar{v}_b$ and $\phi(1) = F^{-1}(0) - G^{-1}(1) = -\bar{v}_s$. Also, $\phi$ is continuous and strictly decreasing in $i$. Therefore, letting $i^*$ denote the unique root of $\phi$, the set of $i \in [0, 1]$ such that $\phi(i) \geq 0$ is the set $[0, i^*]$. Hence, $T = i^*$ and so $\phi(T) = 0$, or $F^{-1}(1-T) = G^{-1}(T)$. Using condition 1, this finally yields $v_b(T) = v_s(T)$.

By condition 3, we see that $v_b(T) \geq p(T) \geq v_s(T)$. However, $v_b(T) = v_s(T)$. Therefore, $p(T) = v_b(T) = v_s(T)$.
Finally, invert condition 1 to get $F[v_B(T)] = 1-T$ and $G[p(T)] = T$. Using $p(T) = v_b(T) = v_s(T)$, we infer that $F[p(T)] = 1-T$ and $G[p(T)] = T$, and so
\begin{equation}G[p(T)] = 1 - F[p(T)] \end{equation}
But this is precisely the equation whose unique solution is $p^*!$ Hence, $p(T) = p^*$.\end{proof}

\newpage

\section{Additional tables and figures} \label{tables_figures}

\setcounter{table}{0}
\renewcommand{\thetable}{B\arabic{table}}

\setcounter{figure}{0}
\renewcommand{\thefigure}{B\arabic{figure}}

\begin{table}[H]
\centering
\caption{Overview of the experimental sessions}\label{overview}
\begin{threeparttable}[h]
\begin{tabular}{cclcc}
\hline
Session & Queue? & \hspace{3em}Treatments       & Rounds & Participants \\ \hline
1                & Yes             & Symmetric then low values & 10              & 18                    \\
2                & Yes             & Low values then symmetric & 10              & 18                    \\
3                & No              & Symmetric then low values & 10              & 10                    \\
4                & No              & Low values then symmetric & 10              & 10                    \\
5                & No              & Low values only           & 9               & 10                    \\
6                & No              & Low values only           & 9               & 10                    \\ \hline
\end{tabular}
\begin{tablenotes}
\footnotesize
\item \hspace{-0.2em}\textit{Notes}: This table describes the differences between our experimental treatments. The second column specifies whether a treatment used a queue of buyers and sellers, and the third column specifies which treatments were run. The fourth and fifth columns specify the number of rounds and participants in the session.
\end{tablenotes}
\end{threeparttable}
\end{table}

\begin{figure}
\centering

\begin{subfigure}{Panel A (Session 5)}
\centering
\includegraphics[width=\linewidth]{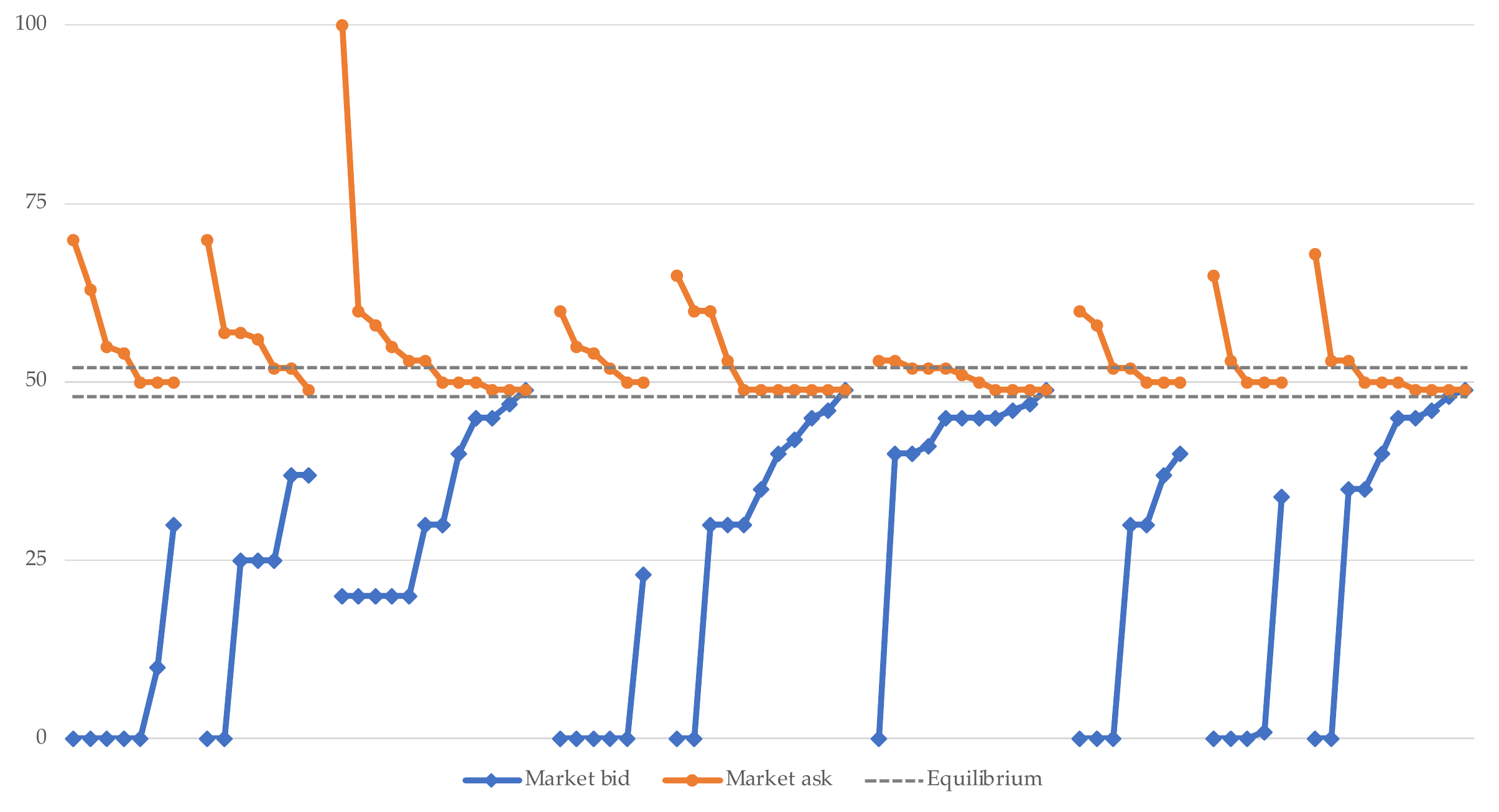}
\end{subfigure}
\begin{subfigure}{Panel B (Session 6)}
\centering
\includegraphics[width=\linewidth]{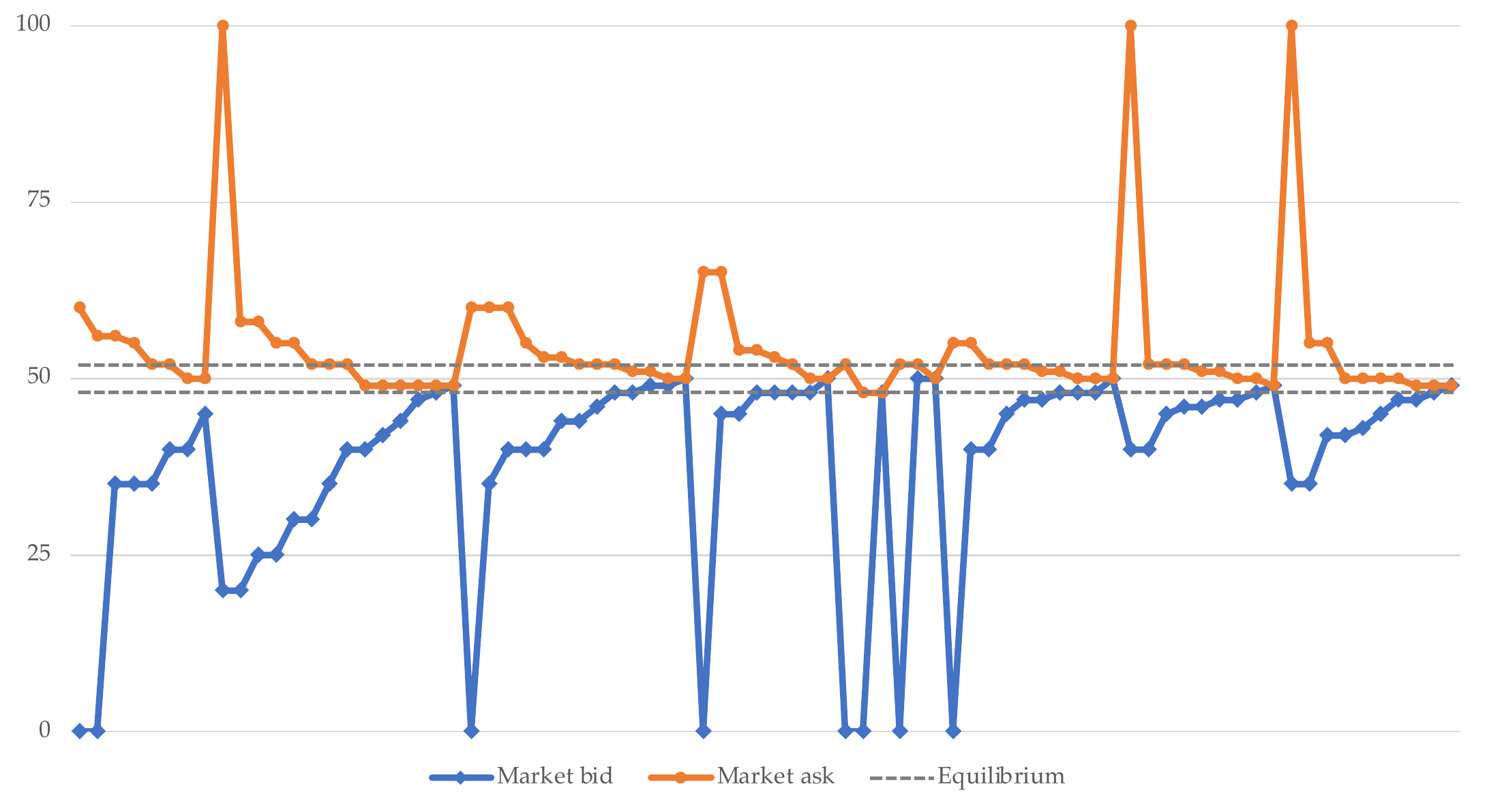}
\end{subfigure}

\medskip

\begin{minipage}[t]{0.2\textwidth}
\caption{Market bids and asks}\label{bids_asks}
\end{minipage}

\begin{minipage}{17cm}%
    \vspace{-3em}\footnotesize \textit{Notes}: This figure reports the market ask and bid in negotiation sequences following a second trade. For simplicity, we discard market asks above £100 and set the market ask at £100 if no asks have yet been made. Similarly, we set the market bid at £0 in the absence of any bids. When a buyer accepts the market ask, we set the market bid equal to the market ask (and likewise for seller acceptances of the market bid). Thus, trade occurs when the market bid and ask meet. The end of a round is denoted by a break.

    \end{minipage}%

\end{figure}

\newpage

\section{Experimental instructions} \label{instructions}

\subsection{Instruction for Buyers (Sessions 1 and 2)}

Welcome to the experiment! Please read the following instructions as carefully as possible.

\textit{Preliminaries}

\begin{itemize}
    \item Please do not talk to your fellow participants at any stage. Talking may result in a loss of experimental earnings.
    \item You will have just received an ID card specifying your buyer number. B1 means `Buyer 1', B2 means ‘Buyer 2’, and so forth. Please wear your ID card visibly at all times.
\end{itemize}

\textit{Roles}

\begin{itemize}
    \item There are two types of buyers in this experiment: active buyers and pending buyers. If your buyer number is between 1 and 5 inclusive, then you are an active buyer and will be sitting in the main trading area. On the other hand, if your buyer number is 6 or higher, then you are a pending buyer and will be sitting in the queue.
    \item Active buyers are free to trade from the very start of a trading period. However, pending buyers may only trade after active buyers have made a trade and dropped out of the market (see elaboration below).
\end{itemize}

\textit{Valuations}

\begin{itemize}
    \item If you are an active buyer, you will have a card marked ‘Valuation’ in front of you.  Examine the number on the back of this card, being careful not to let any other participants see this number. This number is your ‘valuation’ for the fictitious commodity to be traded. Memorise your valuation, turn your ID card back face-down (so the valuation is hidden), and do not reveal your valuation to anybody else!
    \item If you are a pending buyer, then you have not yet been assigned a valuation. However, you will acquire a valuation as soon as an active buyer has dropped out and you have taken that buyer’s place (and valuation).
    \item If you manage to make a trade, you will receive your valuation minus the price that you agreed to pay (this assumes that the trade ‘counts’ — see discussion below).  For example, if your valuation is £55 and you agree to a price of £20, your net earnings are £35.
    \item Notice that your valuation represents the most that you should be conceivably willing to pay to make a trade. For example, if your valuation is £40, then you should never pay more than £40 to buy a unit: doing so would just lose you money!
\end{itemize}

\textit{Trading}

\begin{itemize}
    \item At any point, active buyers may offer to buy at a particular price – this is called making a ‘bid’. To make a bid, raise your hand. Once the auctioneer has pointed at you, state your identity along with how much you want to bid. For example, if you are buyer 3 and you want to bid £40, say ‘buyer 3 bids 40’.
    \item In the sequence of market activity leading up to a trade, each bid must be higher than the current bid. For example, if one buyer bids £20, then all subsequent buyers need to bid more than £20.
    \item All bids need to be whole numbers. For example, while you can bid £30, you cannot bid £30.14.
    \item Just as active buyers may make bids (at any point in time), active sellers can make ‘asks’ (at any point in time). For example, if a seller ‘asks’ for £70, that means that she is willing to sell for £70. Each new ask must be lower than the current ask, so asks must ‘improve’ over time.
    \item At any stage, active buyers may accept an ask that has been made by a seller. To do this, raise your hand and wait until the auctioneer points at you. Once this has occurred, state your identity and that you want to accept the current ask. For example, if you are buyer 2 and want to accept an ask of £70, say ‘buyer 2 accepts 70’.
    \item If multiple buyers want to make a bid or accept an ask, priority will be given to the buyer who has raised their hand first.
    \item If you have made a trade, you become inactive and cannot make any further trades in that round. At this point, you should move to the inactive area and allow your place to be taken by the buyer at the front of the queue (or go to the back of the queue if you were previously a pending buyer). That new buyer will acquire your valuation (so should examine the card on the desk to see what that valuation is).
    \item While each bid needs to be higher than the previous bid, everything is reset following a trade. In other words, once a trade has been made, active buyers are free to submit any bid that they choose – even if that bid is lower than previously submitted bids.
\end{itemize}

\textit{Example}

\begin{itemize}
    \item Buyer 3 wants to bid £40 so raises her hand. Once she is pointed at by the auctioneer, she says ‘buyer 3 bids 40’.
    \item Buyer 1 wants to outbid her, so raises his hand. Once he is pointed at by the auctioneer, he says ‘buyer 1 bids 43’.
    \item Seller 4 wants to offer to sell for £95, so raises her hand. Once she is pointed at by the auctioneer, she says ‘seller 4 asks 95’.
    \item Buyer 2 wants to accept the ask, so raises his hand. Once he is pointed at by the auctioneer, he says ‘buyer 2 accepts 95’. A trade has occurred (at a price of £95).
    \item Since buyer 2 has just traded with seller 4, they both become inactive and should move to the inactive area. Two traders from the queue take their place and acquire their valuation and cost respectively. Participants then continue to bargain over prices. Since a trade has just occurred, subsequent bids are no longer constrained to be above £43 (the previous leading bid).
\end{itemize}

\textit{Rounds}

\begin{itemize}
    \item Trade will continue in this fashion until the auctioneer chooses to end the round.
    \item Within each round, you are only allowed to purchase (at most) one unit of the commodity. However, each round starts afresh: so even if you have made a trade in a particular round, you are free to make trades in subsequent rounds.
    \item After several rounds of trading have concluded, you will be given a new valuation for the fictitious commodity. So even if your current valuation is rather low, you might be luckier later on!
    \item At the end of the experiment, one round will be randomly selected to ‘count’ for calculating your earnings. Since any of the rounds could turn out to be the one that counts, you should do your best to maximise your net earnings within each round.
\end{itemize}

\textit{Summary}

\begin{itemize}
    \item There are two types of buyers: active buyers (IDs 1-5) and pending buyers (IDs 6 or higher). Once active buyers have made a trade, the pending buyer at the front of the queue takes their place and their valuation.
    \item Active buyers may make bids or accept asks at any point in time. Similarly, active sellers may make asks or accept bids at any point in time.
    \item If you want to make a bid or accept an ask, you need to first raise your hand.
    \item Bids must be whole numbers; and new bids must be greater than previous bids (until a trade is made).
    \item If you manage to make a trade, you earn your valuation minus the price you agreed to pay (assuming that this trade is selected to ‘count’ for your earnings).
    \item You can purchase up to one unit within every round.
\end{itemize}

\subsection{Instructions for Sellers (Sessions 1 and 2)}

Welcome to the experiment! Please read the following instructions as carefully as possible.

\textit{Preliminaries}

\begin{itemize}
    \item Please do not talk to your fellow participants at any stage. Talking may result in a loss of experimental earnings.
    \item You will have just received an ID card specifying your seller number. S1 means ‘Seller 1’, S2 means ‘Seller 2’, and so forth. Please wear your ID card visibly at all times.
\end{itemize}

\textit{Roles}

\begin{itemize}
    \item There are two types of sellers in this experiment: active sellers and pending sellers. If your seller number is between 1 and 5 inclusive, then you are an active seller and will be sitting in the main trading area. On the other hand, if your seller number is 6 or higher, then you are a pending seller and will be sitting in the queue.
    \item Active sellers are free to trade from the very start of a trading period. However, pending sellers may only trade after active sellers have made a trade and dropped out of the market (see elaboration below).
\end{itemize}

\textit{Costs}

\begin{itemize}
    \item If you are an active seller, you will have a card marked ‘Cost’ in front of you.  Examine the number on the back of this card, being careful not to let any other participants see this number. This number is how much it would cost you to produce and sell a unit of the fictitious commodity to be traded. Memorise your cost, turn your ID card back face-down (so the cost is hidden), and do not reveal your cost to anybody else!
    \item If you are a pending seller, then you have not yet been assigned a cost. However, you will acquire a cost as soon as an active seller has dropped out and you have taken that seller’s place (and cost).
    \item If you manage to make a trade, you will receive the price paid by the buyer minus your cost (this assumes that the trade ‘counts’ — see discussion below).  For example, if you sell for a price of £55 and your cost is £20, your net earnings are £35.
    \item Notice that your cost represents the least that you should be conceivably willing to sell for. For example, if your cost is £40, then you should never sell for less than £40: doing so would just lose you money!
\end{itemize}

\textit{Trading}

\begin{itemize}
    \item At any point, active sellers may offer to sell at a particular price – this is called making an ‘ask’. To make an ask, raise your hand. Once the auctioneer has pointed at you, state your identity along with how much you are asking for. For example, if you are seller 3 and you want to ask for £60, say ‘seller 3 asks 60’.
    \item In the sequence of market activity leading up to a trade, each ask must be lower than the current ask. For example, if one seller asks £80, then all subsequent sellers need to ask for less than £80.
    \item All asks need to be whole numbers. For example, while you can ask £70, you cannot ask £70.14.
    \item Just as active sellers may make asks (at any point in time), active buyers can make bids (at any point in time). For example, if a buyer bids £70, that means that she is willing to pay £70. Each new bid must be higher than the current bid, so bids must ‘improve’ over time.
    \item At any stage, active sellers may accept a bid that has been made by a buyer. To do this, raise your hand and wait until the auctioneer points at you. Once this has occurred, state your identity and that you want to accept the current bid. For example, if you are seller 2 and want to accept a bid of £70, say ‘seller 2 accepts 70’.
    \item If multiple sellers want to make an ask or accept a bid, priority will be given to the seller who has raised their hand first.
    \item If you have made a trade, you become inactive and cannot make any further trades in that round. At this point, you should move to the inactive area and allow your place to be taken by the seller at the front of the queue (or go to the back of the queue if you were previously a pending seller). That new seller will acquire your cost (so should examine the card on the desk to see what that cost is).
    \item While each ask needs to be lower than the previous ask, everything is reset following a trade. In other words, once a trade has been made, you are free to submit any ask that you choose – even if that ask is higher than previously submitted asks.
\end{itemize}

\textit{Example}

\begin{itemize}
    \item Seller 3 wants to ask for £60 so raises her hand. Once she is pointed at by the auctioneer, she says ‘seller 3 asks 60’.
    \item Seller 1 wants to undercut her, so raises his hand. Once he is pointed at by the auctioneer, he says ‘seller 1 asks 57’.
    \item Buyer 4 wants to bid £5, so raises her hand. Once she is pointed at by the auctioneer, she says ‘buyer 4 bids 5’.
    \item Seller 2 wants to accept the bid, so raises his hand. Once he is pointed at by the auctioneer, he says ‘seller 2 accepts 5’. A trade has occurred (at a price of £5).
    \item Since seller 2 has just traded with buyer 4, they both become inactive and should move to the inactive area. Two traders from the queue take their place and acquire their valuation and cost respectively. Participants then continue to bargain over prices. Since a trade has just occurred, subsequent asks are no longer constrained to be below £57 (the previous lowest).
\end{itemize}

\textit{Rounds}

\begin{itemize}
    \item Trade will continue in this fashion until the auctioneer chooses to end the round.
    \item Within each round, you are only allowed to sell (at most) one unit of the commodity. However, each round starts afresh: so even if you have made a trade in a particular round, you are free to make trades in subsequent rounds.
    \item After several rounds of trading have concluded, you will be given a new cost for the fictitious commodity. So even if your current cost is rather high, you might be luckier later on!
    \item At the end of the experiment, one round will be randomly selected to ‘count’ for calculating your earnings. Since any of the rounds could turn out to be the one that counts, you should do your best to maximise your net earnings within each round.
\end{itemize}

\textit{Summary}

\begin{itemize}
    \item There are two types of sellers: active sellers (IDs 1-5) and pending sellers (IDs 6 or higher). Once active sellers have made a trade, the pending seller at the front of the queue takes their place and their cost.
    \item Active sellers may make asks or accept bids at any point in time. Similarly, active buyers may make bids or accept asks at any point in time.
    \item If you want to make an ask or to accept a bid, you need to first raise your hand.
    \item Asks must be whole numbers; and new asks must be lower than previous asks (until a trade is made).
    \item If you manage to make a trade, you earn the price paid by the buyer minus your cost (assuming that this trade is selected to ‘count’ for your earnings).
    \item You can sell up to one unit within every round.
\end{itemize}

\subsection{Instructions for Buyers (Sessions 3 -- 6)}

Welcome to the experiment! Please read the following instructions as carefully as possible.

\textit{Preliminaries}

\begin{itemize}
    \item Please do not talk to your fellow participants at any stage. Talking may result in a loss of experimental earnings.
    \item You will have just received an ID card specifying your buyer number. B1 means ‘Buyer 1’, B2 means ‘Buyer 2’, and so forth. Please keep your ID card visible at all times.
\end{itemize}

\textit{Valuations}

\begin{itemize}
    \item Please examine the number of the back of your ID card, being careful not to let any other participants see this number. This number is your ‘valuation’ for the fictitious commodity to be traded. Memorise your valuation, turn your ID card back face-down (so the valuation is hidden), and do not reveal your valuation to anybody else!
    \item If you manage to make a trade, you will receive your valuation minus the price that you agreed to pay (this assumes that the trade ‘counts’ — see discussion below).  For example, if your valuation is £55 and you agree to a price of £20, your net earnings are £35.
    \item Notice that your valuation represents the most that you should be conceivably willing to pay to make a trade. For example, if your valuation is £40, then you should never pay more than £40 to buy a unit: doing so would just lose you money!
\end{itemize}

\textit{Trading}

\begin{itemize}
    \item At any point, you may offer to buy at a particular price – this is called making a ‘bid’. To make a bid, raise your hand. Once the auctioneer has pointed at you, state your identity along with how much you want to bid. For example, if you are buyer 3 and you want to bid £40, say ‘buyer 3 bids 40’.
    \item In the sequence of market activity leading up to a trade, each bid must be higher than the current bid. For example, if one buyer bids £20, then all subsequent buyers need to bid more than £20.
    \item All bids need to be whole numbers. For example, while you can bid £30, you cannot bid £30.14.
    \item Just as buyers may make bids (at any point in time), sellers can make ‘asks’ (at any point in time). For example, if a seller ‘asks’ for £70, that means that she is willing to sell for £70. Each new ask must be lower than the current ask, so asks must ‘improve’ over time.
    \item At any stage, you may accept an ask that has been made by a seller. To do this, raise your hand and wait until the auctioneer points at you. Once this has occurred, state your identity and that you want to accept the current ask. For example, if you are buyer 2 and want to accept an ask of £70, say ‘buyer 2 accepts 70’.
    \item If multiple buyers want to make a bid or accept an ask, priority will be given to the buyer who has raised their hand first.
    \item Once you have made a trade, you become inactive and cannot make any further trades in that round. Likewise, each seller is only able to make at most one trade within a round – so it is as if they possess just one unit of the fictitious commodity.
    \item While each bid needs to be higher than the previous bid, everything is reset following a trade. In other words, once a trade has been made, you are free to submit any bid that you choose – even if that bid is lower than previously submitted bids.
\end{itemize}

\textit{Example}

\begin{itemize}
    \item Buyer 3 wants to bid £40 so raises her hand. Once she is pointed at by the auctioneer, she says ‘buyer 3 bids 40’.
    \item Buyer 1 wants to outbid her, so raises his hand. Once he is pointed at by the auctioneer, he says ‘buyer 1 bids 43’.
    \item Seller 4 wants to offer to sell for £95, so raises her hand. Once she is pointed at by the auctioneer, she says ‘seller 4 asks 95’.
    \item Buyer 2 wants to accept the ask, so raises his hand. Once he is pointed at by the auctioneer, he says ‘buyer 2 accepts 95’. A trade has occurred (at a price of £95).
    \item Since buyer 2 has just traded with seller 4, they both become inactive and remain silent for the rest of the round. Meanwhile, other participants continue to bargain over prices. Since a trade has just occurred, subsequent bids are no longer constrained to be above £43 (the previous leading bid).
\end{itemize}

\textit{Rounds}

\begin{itemize}
    \item Trade will continue in this fashion until there is a pause in market activity for roughly 20 seconds. At that point, the auctioneer will ask buyers if they would like to make any new bids, or would like to accept the current market ask. The auctioneer will then ask sellers if they would like to make any new asks, or would like to accept the current market bid. If all traders remain silent, then the auctioneer will close the market and conclude the round.
    \item Within each round, you are only allowed to purchase (at most) one unit of the commodity. However, each round starts afresh: so even if you have made a trade in a particular round, you are free to make trades in subsequent rounds.
    \item After several rounds of trading have concluded, you will be given a new valuation for the fictitious commodity. So even if your current valuation is rather low, you might be luckier later on!
    \item At the end of the experiment, one round will be randomly selected to ‘count’ for calculating your earnings. Since any of the rounds could turn out to be the one that counts, you should do your best to maximise your net earnings within each round.
\end{itemize}

\textit{Summary}

\begin{itemize}
    \item There is a group of buyers, and a group of sellers. Buyers may make bids or accept asks at any point in time. Similarly, sellers may make asks or accept bids at any point in time.
    \item If you want to make a bid or accept an ask, you need to first raise your hand.
    \item Bids must be whole numbers; and new bids must be greater than previous bids (until a trade is made).
    \item If you manage to make a trade, you earn your valuation minus the price you agreed to pay (assuming that this trade is selected to ‘count’ for your earnings).
    \item You can purchase up to one unit within every round.
\end{itemize}

\subsection{Instructions for Sellers (Sessions 3 -- 6)}

Welcome to the experiment! Please read the following instructions as carefully as possible.

\textit{Preliminaries}

\begin{itemize}
    \item Please do not talk to your fellow participants at any stage. Talking may result in a loss of experimental earnings.
    \item You will have just received an ID card specifying your seller number. S1 means ‘Seller 1’, S2 means ‘Seller 2’, and so forth. Please keep your ID card visible at all times. 
\end{itemize}

\textit{Costs}

\begin{itemize}
    \item Please examine the number on the back of your ID card, being careful not to let any other participants see this number. This number is how much it would cost you to produce and sell one unit of the fictitious commodity to be traded. Memorise your cost, turn your ID card back face-down (so the cost is hidden), and do not reveal your cost to anybody else!
    \item If you manage to make a trade, you will receive the price paid by the buyer minus your cost (this assumes that the trade ‘counts’ — see discussion below).  For example, if you sell for a price of £55 and your cost is £20, your net earnings are £35.
    \item Notice that your cost represents the least that you should be conceivably willing to sell for. For example, if your cost is £40, then you should never sell for less than £40: doing so would just lose you money!
\end{itemize}

\textit{Trading}

\begin{itemize}
    \item At any point, you may offer to sell at a particular price – this is called making an ‘ask’. To make an ask, raise your hand. Once the auctioneer has pointed at you, state your identity along with how much you are asking for. For example, if you are seller 3 and you want to ask for £60, say ‘seller 3 asks 60’.
    \item In the sequence of market activity leading up to a trade, each ask must be lower than the current ask. For example, if one seller asks £80, then subsequent sellers need to ask for less than £80.
    \item All asks need to be whole numbers. For example, while you can ask for £70, you cannot ask for £70.14.
    \item Just as sellers may make asks (at any point in time), buyers can make bids (at any point in time). For example, if a buyer bids £70, that means that she is willing to pay £70. Each new bid must be higher than the current bid, so bids must ‘improve’ over time.
    \item At any stage, you may accept a bid that has been made by a buyer. To do this, raise your hand and wait until the auctioneer points at you. Once this has occurred, state your identity and that you want to accept the current bid. For example, if you are seller 2 and want to accept a bid of £70, say ‘seller 2 accepts 70’.
    \item If multiple sellers want to make an ask or accept a bid, priority will be given to the seller who has raised their hand first.
    \item Once you have made a trade, you become inactive and cannot make any further trades in that round --- so it is as if you possess just one unit of the fictitious commodity. Likewise, each buyer is only able to make at most one trade within a round.
    \item While each ask needs to be lower than the previous ask, everything is reset following a trade. In other words, once a trade has been made, you are free to submit any ask that you choose – even if that ask is higher than previously submitted asks.
\end{itemize}

\textit{Example}

\begin{itemize}
    \item Seller 3 wants to ask for £60 so raises her hand. Once she is pointed at by the auctioneer, she says ‘seller 3 asks 60’.
    \item Seller 1 wants to undercut her, so raises his hand. Once he is pointed at by the auctioneer, he says ‘seller 1 asks 57’.
    \item Buyer 4 wants to bid £5, so raises her hand. Once she is pointed at by the auctioneer, she says ‘buyer 4 bids 5’.
    \item Seller 2 wants to accept the bid, so raises his hand. Once he is pointed at by the auctioneer, he says ‘seller 2 accepts 5’. A trade has occurred (at a price of £5).
    \item Since seller 2 has just traded with buyer 4, they both become inactive and remain silent for the rest of the round. Meanwhile, other participants continue to bargain over prices. Since a trade has just occurred, subsequent asks are no longer constrained to be below £57 (the previous lowest ask).
    \item
\end{itemize}

\textit{Rounds}

\begin{itemize}
    \item Trade will continue in this fashion until there is a pause in market activity for roughly 20 seconds. At that point, the auctioneer will ask buyers if they would like to make any new bids, or would like to accept the current market ask. The auctioneer will then ask sellers if they would like to make any new asks, or would like to accept the current market bid. If all traders remain silent, then the auctioneer will close the market and conclude the round.
    \item Within each round, you are only allowed to sell (at most) one unit of the commodity. However, each round starts afresh: so even if you have made a trade in a particular round, you are free to make trades in subsequent rounds.
    \item After several rounds of trading have concluded, you will be given a new cost for the fictitious commodity. So even if your current cost is rather high, you might be luckier later on!
    \item At the end of the experiment, one round will be randomly selected to ‘count’ for calculating your earnings. Since any of the rounds could turn out to be the one that counts, you should do your best to maximise your net earnings within each round.
\end{itemize}

\textit{Summary}

\begin{itemize}
    \item There is a group of sellers, and a group of buyers. Sellers may make asks or accept bids at any point in time. Similarly, buyers may make bids or accept asks at any point in time.
    \item If you want to make an ask or to accept a bid, you need to first raise your hand.
    \item Asks must be whole numbers; and new asks must be lower than previous asks (until a trade is made).
    \item If you manage to make a trade, you earn the price paid by the buyer minus your cost (assuming that this trade is selected to ‘count’ for your earnings).
    \item You can sell up to one unit within every round.
\end{itemize}

\end{appendices}

\end{document}